\documentclass[12pt,twoside,a4paper]{article}
\usepackage{tikz}

\usepackage{geometry,color,framed,multirow}
\usepackage{amsmath,amssymb,mathtools,slashed,xcolor,mdframed}
\usepackage{amstext}
\usepackage{graphicx}
\usepackage{verbatim}
\usepackage{bm}
\usepackage{caption, subcaption}
\usepackage[colorlinks=false, linkcolor= blue,linktoc=true]{hyperref}
\makeatletter
\newcommand{\rd}{\hspace{4mm}\Rightarrow\hspace{4mm}}
\newcommand{\mc}{\mathcal}
\newcommand{\beq}{\begin{eqnarray}}
\newcommand{\eeq}{\end{eqnarray}}
\newcommand{\non}{\nonumber\\}
\newcommand{\ben}{\begin{itemize}}
\newcommand{\enn}{\end{itemize}}
\newcommand{\ra}{\rightarrow}
\newcommand{\mb}{\mathbb}

\newcommand{\lf}{\left}
\newcommand{\rr}{\right}
\newcommand{\ttx}{\texttt}
\newcommand{\tb}{\textbf}
\def\pd{{\partial}}
\def\S{{\Sigma}}
\def\m{{\mu}}
\def\n{{\nu}}
\def\a{{\alpha}}
\def\b{{\beta}}
\def\d{{\delta}}

\def\p{{\phi}}
\def\c{{\chi}}
\def\g{{\gamma}}
\def\ep{{\epsilon}}

\input{epsf.sty}  
\begin{document}

\begin{center}
\[\]\[\]\[\]
\LARGE{Cosmological solutions from 4D \\$N=4$ matter-coupled supergravity}
\end{center}
\vspace{1cm}
\begin{center}
\large{H. L. Dao}
\end{center}
\begin{center}
\textsl{Department of Physics,\\
 National University of Singapore,\\
3 Science Drive 2, Singapore 117551}\\
\texttt{hl.dao@nus.edu.sg}\\
\vspace{2cm}
\end{center}
\begin{abstract}
From four-dimensional $N=4$ matter-coupled gauged supergravity, we study smooth time-dependent cosmological solutions interpolating between a $dS_2\times \Sigma_2$ spacetime, with $\Sigma_2 = S^2$ and $H^2$, in the infinite past and a $dS_4$ spacetime in the infinite future. The solutions were obtained by solving the second-order equations of motion from all the ten gauged theories known to admit $dS_4$ solutions, of which there are two types. Type I $dS$ gauged theories can admit  both $dS$ solutions as well as supersymmetric $AdS$ solutions while type II $dS$ gauged theories only admit $dS$ solutions. We also study the extent to which the first-order equations that solve the aforementioned second-order field equations fail to admit the $dS_4$ vacua and their associated cosmological solutions. 
\end{abstract}
\newpage
\setcounter{tocdepth}{2}
\tableofcontents
\section{Introduction}\label{intro}
Cosmological solutions connecting a $dS_4$ spacetime with a very large Hubble constant in the past to another $dS_4$ spacetime with a very small Hubble constant (both of the right order of magnitude) in the future can provide a good description of our universe. In the context of the dS/CFT holography, as formulated in \cite{Strominger-01}, \cite{Strominger-01-2}, these and the more general cosmological solutions connecting $dS$ spacetimes of different dimensions, such as $dS_{D-d}\times \S^d$ and $dS_D$,  are fundamental building blocks on the gravity side. These are dual to Euclidean conformal field theories with different dimensions which are connected by time evolution since the holographic dimension is time itself. Despite much effort, a meta-stable $dS$ vacuum remains elusive from string theory \cite{Danielsson-18}. Furthermore, there exist many outstanding issues regarding the quantum nature of $dS$ space itself \cite{Witten-10}. 
Because of these difficulties, dS/CFT holography based on the correspondence between a $dS_{D+1}$ solution of $(D+1)$-dimensional supergravity and a Euclidean CFT in $D$ dimensions, analogous to the now routinely-applied AdS/CFT correspondence between an $AdS_{(D+1)}$ spacetime and a CFT$_D$ theory, remains very poorly understood. To date, arguably the most concrete form of dS/CFT is found in higher spin theories \cite{Strominger-11}, \cite{Anninos-14}, \cite{Anninos-17} instead of ordinary supergravity. 
\\\indent
A radically different way to obtain dS solutions is from unconventional gauged $dS$ supergravities that arise from the compactification of the exotic 11D M* and 10D type IIB* theories \cite{Liu-Sabra-03}, which are themselves obtained from the timelike T-dualities of the original M/string theories \cite{Hull-98-Jul}, \cite{Hull-98-Aug}, \cite{Hull-99-time}. This framework, originally proposed in \cite{Hull-98-Jun}, naturally gives dS space the priviledged position occupied by AdS space in the conventional framework of M/string theories. Consequently, dS/CFT holography, in the sense proposed by \cite{Hull-98-Jun}, naturally arises in the context of these *-theories in the same manner that AdS/CFT holography does in conventional M/string theories. It is then unsurprising that many cosmological solutions can be found in this framework, see for examples \cite{Lu-03}, \cite{Lu-03-flow}. 
\\\indent
On the other hand, only a few examples involving cosmological solutions in conventional gauged supergravities are known. These include \cite{Cvetic-noncompact} and \cite{Lu-04}. The cosmological solution in \cite{Cvetic-noncompact} interpolates between a $dS$ spacetime and a singularity, while the cosmological solution  in \cite{Lu-04}, interpolating between a $dS_2\times S^{D-2}$ spacetime in the past and a $dS_D$ spacetime in the future, is an analytic continuation of the domain wall solution interpolating between an $AdS_2\times S^{D-2}$ spacetime and an $AdS_D$ spacetime. 
In \cite{dS5-cosmo}, cosmological solutions  in 5D $N=4$ supergravity interpolating between a $dS_{5-d}\times H^d$ (with $d=2,3$) spacetime in the infinite past and a $dS_5$ spacetime in the infinite future were presented in all the gauged theories capable of admitting $dS_5$ solutions. Since the $dS_5$ vacua of 5D $N=4$ supergravity are unstable, these and their associated cosmological vacua studied in \cite{dS5-cosmo} are not relevant in the context of $dS$ holography. Nonetheless, independently of holography, it was instructive to derive and study the cosmological solutions in 5D $N=4$ supergravity in their own right. The main motivation for the study in \cite{dS5-cosmo} was straightforward - we would like to know whether there exist fixed point solutions $dS_{5-d}\times \S_d$, with $\S_d = S^d, H^d$ where $d=2,3$, and interpolating solutions connecting these in the infinite past to the $dS_5$ solutions in the infinite future. 
It is our goal in this work to carry out the same analysis in the four-dimensional, $N=4$ supergravity that is structurally similar to, but more complex than the five-dimensional theory studied in \cite{dS5-cosmo}. More specifically, our primary goal will be to study cosmological solutions interpolating between a $dS_2\times \S_2$ spacetime, with $\S_2 = S^2, H^2$ , at early times and a $dS_4$ spacetime at late times from 4D $N=4$ matter-coupled gauged supergravities. The solutions will be derived by solving the second-order field equations from the ten gauged theories known to admit $dS_4$ solutions. These gauged theories can be classified into two types, depending on whether the gauge group under consideration allows for an $AdS$ solution in addition to a $dS$ solution, or just $dS$ solutions \cite{dS4}. Secondly, similar to the 5D case in \cite{dS5-cosmo}, our secondary goal will be to study the possibility of obtaining these $dS_4$  and  cosmological solutions  from some system of first-order equations that solve the second-order field equations. Although the fact that all $dS_4$ vacua of 4D $N=4$ supergravity are unstable precludes the existence of pseudosupersymmetry which guarantees the existence of first-order pseudo-BPS equations capable of giving rise to these solutions, we will nevertheless perform our analysis, analogous to the 5D case, to pinpoint the exact failure characterizing the lack of first-order systems.
\\\indent
This rest of this paper is organized as follows. In section \ref{4D-N4-review}, we review the theory of 4D $N=4$ gauged supergravity coupled to vector multiplets in the embedding tensor formalism. In section \ref{4D-dS4}, we review the $dS_4$ solutions derived within the framework of the embedding tensor formalism and their classifications into type I and type II.  In section \ref{4D-dS2}, the ansatze for the $dS_2\times \S_2$ solutions are specified and  the associated equations of motion are derived. In section \ref{4D-dS2-t1} and \ref{4D-dS2-t2}, we present the cosmological solutions from the type I and type II $dS$ gauged theories, respectively. In section \ref{BPS}, we investigate how the cosmological solutions found in sections \ref{4D-dS2-t1},\ref{4D-dS2-t2} fail to arise from the first-order equations that solve the second-order field equations. Section \ref{concl} concludes the paper. 
\section{4D $N=4$ matter-coupled supergravity}\label{4D-N4-review}
To make this paper self-contained, we highlight some relevant details of four-dimensional $N=4$ matter-coupled gauged supergravity. The  detailed construction of theory can be found in \cite{Schon-Weidner} and the references therein. 
\\
\indent
The field content of the theory includes an $N=4$ supergravity multiplet and an arbitrary number $n$ of vector multiplets. The supergravity multiplet
\beq
\lf(e^{\hat \mu}_\mu, \,\, \psi_{\m \,i}, \,\, A^m_\m, \,\,\lambda_i,\,\,\tau \rr)\nonumber
\eeq
 contains the graviton $e^{\hat \mu}_\mu$, four gravitini $\psi_{\mu i}$, $i=1,...,4$, six vectors $A^m_\mu$, $m=1,...,6$, four spin-$\frac{1}{2}$ fermions $\lambda_i$, and a complex scalar $\tau$. The vector multiplet
 \beq
 \lf(A_\m, \,\,\lambda^i, \,\,\phi^m \rr) \nonumber
 \eeq
contains a vector field $A_\mu$, four gaugini $\lambda^{i}$ and six scalars $\phi^{m}$. Spacetime and tangent space indices will be denoted by $\mu, \nu, \ldots = 0,1,2,3$ and $\hat\mu, \hat \nu,\ldots = 0,1,2,3$, respectively. Indices $m,n,\ldots=1,2,\ldots, 6$ label vector representation of $SO(6)\sim SU(4)$ R-symmetry while $i,j,\ldots$ denote chiral spinor of $SO(6)$ or fundamental representation of $SU(4)$. The $n$ vector multiplets are labeled by indices $a,b,\ldots =1,...,n$. Collectively, the fied content of the $n$ vector multiplets can be written as 
\beq
(A^a_\mu,\lambda^{ai},\phi^{am}).\nonumber
\eeq 
\indent Altogether, there are $(6n+2)$ scalars from both the gravity and vector multiplets. These scalars span the coset manifold
\begin{eqnarray}
\mathcal M = \frac{SL(2, \mb R)}{SO(2)}\times \frac{SO(6,n)}{SO(6)\times SO(n)}\,.\label{Mscalar}
\end{eqnarray}
The first factor in (\ref{Mscalar}) is parameterized by a complex scalar $\tau$ consisting of the dilaton $\phi$ and the axion $\chi$ from the gravity multiplet, where
\beq
\tau = \chi + i \, e^\phi\,\,.
\eeq
 The second factor in (\ref{Mscalar}) is parameterized by the $6n$ scalars from the vector multiplets. Both factors of (\ref{Mscalar}) will be described in terms of coset representatives. 
For $SL(2)/SO(2)$, this coset representative is
\begin{eqnarray}
\mc V_\alpha = \frac{1}{\sqrt{\textrm{Im}\tau}} \begin{pmatrix}\tau \\ 1\end{pmatrix}
\end{eqnarray}
with an index $\alpha = (+,-)$ denoting the $SL(2)$ fundamental representation. In the next section, we will also be using the notation $\a = (e,m)$. 
$\mc{V}_\alpha$ satisfies the relation
\begin{equation}
M_{\alpha\beta}=\textrm{Re}(\mc{V}_\alpha\mc{V}^*_\beta)\qquad \textrm{and}\qquad \epsilon_{\alpha\beta}=\textrm{Im}(\mc{V}_\alpha\mc{V}^*_\beta)
\end{equation}
in which $M_{\alpha\beta}$ is a symmetric matrix with unit determinant, while $\epsilon_{\alpha\beta}$ is anti-symmetric with $\epsilon_{+-}=\epsilon^{+-}=1$. 
\\
\indent The coset manifold $SO(6,n)/SO(6)\times SO(n)$ will be described by a coset representative ${\mc {V}_M}^A$ transforming under the global $SO(6,n)$ and local $SO(6)\times SO(n)$ symmetry by left and right multiplications, respectively. The local index $A$ can be split as $A=(m,a)$ with $m=1,2,\ldots,6$ and $a=1,2,\ldots ,n$ denoting vector representations of $SO(6)\times SO(n)$. Accordingly, ${\mc V_M}^A$ can be written as
\begin{eqnarray}
{\mc V_M}^A = ({\mc V_M}^m, {\mc V_M}^a)\, .
\end{eqnarray}
With ${\mc V_M}^A$ being an $SO(6,n)$ matrix, the following relation is satisfied
\begin{eqnarray}
\eta_{MN} = -{\mc V_M}^m {\mc V_N}^m + {\mc V_M}^a {\mc V_N}^a
\end{eqnarray}
where $\eta_{MN} = \textrm{diag}\left(-1,-1,-1,-1,-1,-1, 1,\ldots,1\right)$ is the $SO(6,n)$ invariant tensor. $SO(6,n)$ indices $M, N, \ldots$ are lowered and raised using $\eta_{MN}$ and its inverse $\eta_{MN}$.
Equivalently, the symmetric matrix 
\begin{equation}
M_{MN} ={\mc V_M}^m {\mc V_N}^m + {\mc V_M}^a {\mc V_N}^a,
\end{equation}
which is manifestly $SO(6)\times SO(n)$ invariant, is used to describe the $SO(6,n)/SO(6)\times SO(n)$ coset.
\\
\indent Gaugings of the matter-coupled $N=4$ supergravity are performed using the embedding tensor formalism \cite{Schon-Weidner} that provides an embedding of a gauge group $G_0$ in the global symmetry group $SL(2,\mb R)\times SO(6,n)$. $N=4$ supersymmetry restricts the embedding tensor to include only the components  $\xi_{\alpha M}$ and  $f_{\alpha MNP}=f_{\alpha[MNP]}$. A closed subalgebra of $SL(2, \mb R)\times SO(6,n)$ satisfying the required commutation relations is ensured by the following quadratic constraints on the components of the embedding tensor
\begin{eqnarray}
0 &=& {\xi_\alpha}^M\xi_{\beta M},\non
 0 &=&\epsilon^{\alpha\beta}({\xi_\alpha}^Pf_{\beta_{MNP}}+\xi_{\alpha M}\xi_{\beta N}),\non
0 &=&{\xi_{(\alpha}}^Pf_{\beta) MNP},\non
0 &=&3f_{\alpha R[MN}{f_{\beta PQ]}}^R+2\xi_{(\alpha [M}f_{\beta)NPQ]},
\nonumber \\
0 &=&\epsilon^{\alpha\beta}\left(f_{\alpha MNR}{f_{\beta PQ}}^R-{\xi_\alpha}^Rf_{\beta R[M[P}\eta_{Q]N}-\xi_{\alpha [M}f_{\beta N]PQ}+\xi_{\alpha [P}f_{\beta Q]MN}\right)\, .\label{QC-4D}
\end{eqnarray}
The gauge group generators are constructed from these components of the embedding tensor as follows. 
\beq
{X_{\a M,\,\b N}}^{\gamma P} = -\d_\b^\gamma {f_{\a MN}}^P + \frac{1}{2}\lf[ \d^P_M \d_\b^\gamma \,\xi_{\a N} - \d_N^P \d_\a^\g \,\xi_{\b N} - \d_\b^\g \eta_{MN} \xi_\a^P + \epsilon_{\a\b} \d_N^P \xi_{\kappa M} \ep^{\kappa\g}\rr]
\label{G-gen}
\eeq
In this work, we are only interested in solutions involving the metric, scalars and some Abelian gauge fields.
The bosonic Lagrangian reads
\beq
e^{-1}\mc L &=& \frac{1}{2}R + \frac{1}{16}D_\m M_{MN} D^\m M^{MN} + \frac{1}{8} D_\m M_{\a\b} D^\m M^{\a\b} - V\non
&&
-\frac{1}{4}\text{Im}\,\tau\,M_{MN}\mc H^{M+}_{\m\n}\mc H^{N+\m\n} + \frac{1}{8}\,\text{Re}(\tau) \,\eta_{MN}\ep^{\m\n\rho\lambda}\mc H^{M+}_{\m\n} H_{\rho\lambda}^{N+} + e^{-1}\mc L_\text{top}
 \label{L-full}
\eeq
 where $e$ is the vielbein determinant, and $\mc L_\text{top}$ is the topological term whose explicit form will not be of relevance to our purpose because this term will always vanish for the solutions considered here. Note that in the Lagrangian above, only covariant electric gauge field strengths $\mc H^{M+}_{\m\n}$ are present while magnetic gauge fields $\mc H^{M-}_{\m\n}$ enter the equations of motion. The covariant derivatives acting on the scalars\footnote{Note that there are two equivalent ways to write the kinetic term for the supergravity scalar $\tau$
 \beq
 D_\m M_{\a\b}\, D^\m M^{\a\b}  = -\frac{2}{\text{Im}(\tau)^2} D_\m \tau \,D^\m \tau^*.
 \eeq} read
 \beq
 D_\m M_{MN} &=& \pd_\m M_{MN} + 2{A_\m}^{P\a}{\Theta_{\a P(M}}^Q M_{N)Q}\non
 D_\m M_{\a\b} &=& \pd_\m M_{\a\b} + A_\m^{M\g}\xi_{(\a M} M_{\b)\g} - A_\m^{M\d}\xi_{\kappa M}\epsilon_{\d(\a}\epsilon^{\kappa\gamma}M_{\b)\g}\,\,.
 \eeq
 where
 \beq
 \Theta_{\a MNP} = f_{\a MNP} - \xi_{\a[N}\eta_{P]M}.
 \eeq
 \indent In general, the covariant field strengths $\mc H^{M\a}_{\m\n}$ contain the auxiliary two-form fields $B^{[MN]}_{\m\n}$ and $B^{(\a\b)}_{\m\n}$ that transform in the antisymmetric representation of $SO(6,n)$ and in the symmetric representation of $SL(2, \mb R)$, respectively. These two-forms are needed to preserve the covariance of the gauge field strengths $\mc H^{M\a}_{\m\n}$. 
 At this point we refrain from giving the full form of $\mc H^{M\a}_{\m\n}$ that contain $B^{MN}_{\m\n}$ and $B^{\a\b}_{\m\n}$ due to their complexity and the fact that these two-forms can be consistently truncated out for the specific embedding tensor components corresponding to all the gauge groups studied in this work. With $B^{[MN]}_{\m\n} = B^{\a\b}_{\m\n} = 0$ and the fact that only Abelian gauge fields are turned on, the gauge field strengths read
 \beq
 \mc H^{M\a}_{\m\n} = 2\pd_{[\m} A^{M\a}_{\n]}\,\,.
 \eeq
The scalar potential $V$ reads 
\begin{eqnarray}
V&=&\frac{g^2}{16}\left[f_{\alpha MNP}f_{\beta
QRS}M^{\alpha\beta}\left[\frac{1}{3}M^{MQ}M^{NR}M^{PS}+\left(\frac{2}{3}\eta^{MQ}
-M^{MQ}\right)\eta^{NR}\eta^{PS}\right]\right.\nonumber \\
& &\left.-\frac{4}{9}f_{\alpha MNP}f_{\beta
QRS}\epsilon^{\alpha\beta}M^{MNPQRS}+3{\xi_\alpha}^M{\xi_\beta}^NM^{\alpha\beta}M_{MN}\right] \label{Vs}
\end{eqnarray}
where $M^{MN}$ and $M^{\alpha\beta}$ are the inverse matrices of $M_{MN}$ and $M_{\alpha\beta}$, respectively. $M^{MNPQRS}$ is obtained by raising indices of $M_{MNPQRS}$ defined by
\begin{equation}
M_{MNPQRS}=\epsilon_{mnpqrs}\mc{V}_{M}^{\phantom{M}m}\mc{V}_{N}^{\phantom{M}n}
\mc{V}_{P}^{\phantom{M}p}\mc{V}_{Q}^{\phantom{M}q}\mc{V}_{R}^{\phantom{M}r}\mc{V}_{S}^{\phantom{M}s}\, .\label{M_6}
\end{equation}
\indent Alternatively, the potential $V$ can be written in terms of the fermion-shift matrices $A_1^{ij}$, $A_2^{ij}$ and ${A_{2ai}}^j$ that appear in the fermion mass-like terms and supersymmetry transformations of fermions as follows. 
\begin{eqnarray}
V = \frac{1}{2}{A_{2ai}}^j {A^*_{2aj}}^i + \frac{1}{9}A_2^{ij}A^*_{2ij} -\frac{1}{3}A^{ij}_1 A^*_{1ij}\, . \label{V-A}
\end{eqnarray}
In terms of the scalar coset representatives, the fermion shift-matrices are given by
\begin{eqnarray}
A^{ij}_1 &=& \epsilon^{\alpha\beta}(\mc V_\alpha)^* {\mc {V}_{kl}}^M {\mc {V}_N}^{ik}{\mc V_P}^{jl} {f_{\beta M}}^{NP}, \label{A1} \\
A^{ij}_2 &=& \epsilon^{\alpha\beta} \mc V_\alpha {\mc {V}_{kl}}^M {\mc V_N}^{ik} {\mc V_P}^{jl} {f_{\beta M}}^{NP}+\frac{3}{2}\epsilon^{\alpha\beta}\mc{V}_\alpha{\mc{V}_M}^{ij}{\xi_\beta}^M, \label{A2}\\
{A_{2ai}}^j &=& \epsilon^{\alpha\beta} \mc V_\alpha {\mc V_a}^M {\mc V_{ik}}^N {\mc V_P}^{jk} {f_{\beta MN}}^{P}-\frac{1}{4}\delta^j_i\epsilon^{\alpha\beta}\mc{V}_\alpha {\mc{V}_a}^M\xi_{\beta M}\, .\label{A2a}
\end{eqnarray}
${\mc{V}_M}^{ij}$ is obtained from ${\mc{V}_M}^m$ by converting the $SO(6)$ vector index to an anti-symmetric pair of $SU(4)$ fundamental indices using the chiral $SO(6)$ gamma matrices. The explicit gamma matrices used are given in the Appendix of \cite{dS4}.
\section{$dS_4$ solutions from 4D $N=4$ supergravity}\label{4D-dS4}
In this section we summarize the known results on $dS_4$ solutions from \cite{dS4}. 
$dS_4$ solutions of 4D $N=4$ gauged supergravity coupled to $n$ vector multiplets were originally derived in \cite{deRoo-Panda1}, \cite{deRoo-Panda2}, and rederived within the framework of the embedding tensor formalism in \cite{dS4}
 using the following two ansatze for the fermionic shift matrices (\ref{A1}, \ref{A2}, \ref{A2a})
\beq
\langle A_1^{ij}\rangle = \langle {A_{2ai}}^j\rangle = 0, \qquad \langle A_2^{ik} A^*_{2jk}\rangle = \frac{9}{4} V_0  \d^i_j
\label{ans-1}
\\
\langle A_1^{ij}\rangle = \langle {A_{2}}^{ij}\rangle = 0, \qquad \langle {A_{2ai}}^{k} {A^*_{2ak}}^j\rangle = \frac{1}{2} V_0  \d^i_j
\label{ans-2}
\eeq
where $V_0$ is the extremized value of the potential, and $\langle \,\,\,\rangle$ denotes the enclosed quantities being evaluated in the considered background. We recall that all solutions from \cite{deRoo-Panda1} and \cite{deRoo-Panda2} were recovered in \cite{dS4} and in addition, a new solution with the gauge group $SO(2,1)\times SO(4,1)$ was found in \cite{dS4}.
While the ansatze (\ref{ans-1}, \ref{ans-2}) ensure the positivity of the scalar potential, they must also be subject to the   quadratic constraints (\ref{QC-4D}), and the extremization condition on the scalar potential. 
For more details, see \cite{dS4}. Solutions to either (\ref{ans-1}) or (\ref{ans-2}) consist of the set of embedding tensor components ($\xi_{\a\,M},\, f_{\a\,MNP}$) from which the gauge group generators can be constructed using (\ref{G-gen}). As shown in \cite{dS4}, there exist solutions to both  (\ref{ans-1}) and (\ref{ans-2}) which form the two types of gaugings that give rise to two different types of $dS$ solutions in gauged $N=4$ four dimensional supergravity.
\\\indent
The common features of these two types are as follows. 
 Firstly, all gaugings with $dS_4$ solutions must have $\xi_{\alpha  M} = 0$ which simplifies the formula for the gauge generators (\ref{G-gen})
 \beq
 {X_{\a M\, N}}^{\, P} = -{f_{\a\,M N}}^P. \label{G-gen-2}
 \eeq
 Secondly, the gauge groups $G_0$ must be dyonic, comprising a product  of at least one electric $G_e$ and one magnetic $G_m$ gauge factor under the $SL(2,\mb R)$ duality group\footnote{Here we use the notation $\a = (e,m)$ instead of $\a = (+,-)$ as used in section \ref{4D-N4-review} to avoid the potential confusion arising from the use of subscript $(+,-)$ reserved for other meanings.}.
 \beq
 G_0 = G_e\times G_m.
 \eeq
 The differences between these two types of gaugings are as follows.  
\begin{itemize}
	\item \textit{Type I $dS_4$ solutions} arise from gaugings of the first type, which can admit both $dS$ and $AdS$ solutions.  The compulsory components of the 4D embedding tensor in this case are
	\beq
	f_{\alpha\,m n p} \neq 0. \label{t1-c}
	\eeq
	Other allowed (but not required) components of the embedding tensors are
	\beq
	f_{\alpha\, m  a b} \neq 0, \qquad f_{\alpha \, a  b  c} \neq 0.
	\eeq
	The gauge group is of the form 
	\beq 
	G_0 = G_{e, -}\times G_{m, -}\times \ldots \label{G0-t1}
	\eeq
	where $G_{e, -}, G_{m, -},\ldots $ are \textit{in general} non-compact. The compact parts of (\ref{G0-t1}) are always $SO(3)_e\times SO(3)_m$, generated by $f_{\alpha\,m n p}$\footnote{ For $AdS$ vacua, this compulsory $SU(2)_e\times SU(2)_m$ factor can be identified with the R-symmetry group of 3d $\mc N=4$ SCFTs according to the AdS/CFT correspondence.}, and lie entirely in the R-symmetry group $SO(6)\subset SO(6,n)$ of (\ref{Mscalar}).  The non-compact parts of (\ref{G0-t1}), generated by $f_{\alpha\, m  a b}$, lie completely in the matter-symmetry directions $SO(n)\subset SO(6,n)$ of (\ref{Mscalar}). Accordingly, the $-$ subscript in the $G_{e(m)}$ factors in (\ref{G0-t1}) is used to denote the embedding of the compact parts of (\ref{G0-t1}) in the R-symmetry directions. 
	Additionally, $f_{\alpha \, a  b  c}$  generate a purely compact gauge factor, embedded in the matter symmetry directions $SO(n)$, that is optional and does not contribute anything to the scalar potential.  
	 When $f_{\alpha\, m  a b}$ vanish, we can have the completely compact gauging $SO(3)_{e, -}\times SO(3)_{m,-}$ which can exist in the pure supergravity theory without any coupled matter. 
	\item \textit{Type II $dS_4$ solutions}  arise from gaugings of the second type, which can only admit  $dS$ solutions with no $AdS$ solutions possible.  The compulsory components of the 4D embedding tensor in this case are
	\beq
f_{\alpha\,a m  n} \neq 0.\label{t2-c}
	\eeq
		Other allowed (but not required) components of the embedding tensors are
	\beq
	f_{\alpha\, a b c} \neq 0.
	\eeq
The gauge group is \textit{always} noncompact and is of the product form 
	\beq
	G_0 = G_{e, +} \times G_{m,+}\times \ldots \label{G0-t2}
	\eeq
	The non-compact parts of (\ref{G0-t2})  are generated by the components $f_{\alpha\, a   m  n}$ and lie entirely along the R-symmetry directions $SO(6)\subset SO(6,n)$ of (\ref{Mscalar}).  The compact parts of (\ref{G0-t2}) are generated by $f_{\a\, a b c}$ and lie fully in the matter symmetry directions $SO(n)\subset SO(6,n)$ of (\ref{Mscalar}). Accordingly, the $+$ subscript in the $G_{e(m)}$ factors of (\ref{G0-t2}) is used to denote the embedding of the compact parts of (\ref{G0-t2}) in the matter symmetry directions. This is in direct constrast to those gaugings of the first type above.  When $f_{\alpha\, a b c} =0$, the only gauging possible is $SO(2,1)_{e,+}\times SO(2,1)_{m,+}$.  Just like in 5D \cite{dS5}, gaugings of this second type cannot exist in the pure theory, but only in the matter-coupled theories.
\end{itemize}
We emphasize that the $+/-$ subscripts in (\ref{G0-t1}) and (\ref{G0-t2}) are not to be mistaken with the earlier notation of $\a=(+,-)$ used in section \ref{4D-N4-review}. Here the $\a$ subscript is explicitly labeled as $\a = (e,m)$, while the $+$ or $-$ subcripts are used to denote whether the compact parts of the gauge group are embedded in the matter or R-symmetry directions. 
The embedding tensor components for all 4D gaugings with $dS_4$ solutions are listed in table \ref{dS4-sum} where we have dropped the $e(m)$ subscripts, but retained the ($+/-$) subscripts in the gauge factors. Because of the $SL(2,\mb R)$ duality, either  $G_+$ or $G_-$ can be electric or magnetic and vice versa - this is reflected from the embedding tensor components $f_{\a MNP}$ where $\a$ can assume either the value of $e$ or $m$. 
\\\indent
While the full form of all gauge groups is given as in Table \ref{dS4-sum}, we must point out that not all gauge factors in these gauge groups contribute to the scalar potentials, as mentioned above. Only the compulsory components (\ref{t1-c}), (\ref{t2-c}) are needed for each type of gaugings. Consequently, to simplify the analysis, we can reduce some of the gauge groups to the bare forms needed to admit $dS_4$ solutions without any loss of generality. 
 For later convenience, we  summarize the explicit scalar potentials corresponding to the simplified versions of those gauge groups listed in Table \ref{dS4-sum} in Table \ref{dS4-sum-V}. These scalar potentials are constructed from the explicit components of the embedding tensor given in Table \ref{dS4-sum} using (\ref{Vs}) or (\ref{V-A}). For concreteness,  $g_1$ and $g_2$ are used to denote the electric factor $G_e$ and magnetic factor $G_m$, respectively, of the gauge group $G_0$. From here on, to lighten the notations, the explicit $+/-$ subscripts in the gauge groups will also be dropped since their distinction is made clear from their assignment to either type I or type II gaugings. It is important to note that while the scalar potentials for the type II gaugings are all different, there is only a single scalar potential resulting from the type I gaugings. 
\begin{small}
 \begin{table}[htb!]
\begin{center}
\begin{tabular}{|c|cl|}
\hline
& Gauge group $G_0$ & Embedding tensor $f_{\a MNP}$
\\\hline
\multirow{16}{*}{$\begin{array}{c}\text{Type I:}\\\text{$AdS_4$ \& $dS_4$} \\ \text{possible}\end{array}$} &  $SO(3)_+^2\times SO(3)_-^2$ &$\begin{array}{l} f_{\alpha\,123} = g_1, \,\,f_{\alpha\,789} = \tilde g_1, \\f_{\beta 456} = g_2, \,\,f_{\beta 10,11,12} = \tilde g_2\end{array}$ \\&&\\
& $SO(3,1)_-\times SO(3,1)_-$ & $\begin{array}{l} f_{\alpha\,123} = -f_{\alpha \,783} = f_{\alpha\,729} = f_{\alpha\,189} = g_1 \\ f_{\beta\,456} = -f_{\beta\,10,11,12} =g_2\\ f_{\beta 10,5,12} = f_{\beta\,4,11,12} = g_2\end{array}$
 \\&&\\
    & $SO(3)_+\times SO(3)_-\times SO(3,1)_-$ &$\begin{array}{l}   f_{\alpha\,123} = f_{\alpha\,789} =  g_1 \\ f_{\beta\,456} = -f_{\beta\,10,11,6} =g_2\\ f_{\beta\,10,5,12} = f_{\beta\,4,11,12} = g_2 \end{array}$  \\&&\\
& $SO(3)_-\times SL(3,\mathbb R)_-$ &$\begin{array}{l} f_{\alpha\,123} = f_{\alpha\,279}=  f_{\alpha\, 389}  = -g_1\\ f_{\alpha,1,9,10} = -f_{\alpha,2,8,10}= f_{\alpha,3,7,10} =-g_1\\ f_{\alpha\,2,10,11} = f_{\alpha\,3,9,11} = \sqrt{3}g_1\\
f_{\alpha\,178} = 2g_1, \,\,f_{\beta\,456} = g_2\end{array}$ \\
\hline
    \multirow{22}{*}{$\begin{array}{c}\text{Type II: } \\ \text{Only $dS_4$}\\ \text{possible}\end{array}$}& $SO(2,1)^2_+\times SO(2,1)_-^2$   &$\begin{array}{l} f_{\alpha 723} = g_1,\,\,f_{\alpha \,189} = \tilde g_1 \\ f_{\beta\,10,5,6} = g_2,\,\,f_{\beta\,4,11,12} = \tilde g_2\end{array}$ \\&&\\
    & $ SO(2,1)_+\times SO(2,2)_+\times SO(3)_+$ & $\begin{array}{l} f_{\alpha\,569} = g_2, f_{\alpha,10,11,12} = g_3 \\ f_{\beta \,127} =  g_1, \,\,\,f_{\beta\,348} =  g_1\end{array}$ \\ &&\\
        & $SO(2,1)_+\times SO(3,1)_+\times SO(2,1)_-$ &$\begin{array}{l}  f_{\alpha\,789} = -f_{\alpha\,129} = f_{\alpha\,183} = f_{\alpha\,723} = g_1 \\
    f_{\beta\,10,5,6} = g_2, \,\,\,f_{\beta\,4,11,12} = \tilde g_2 \end{array}$ \\&& \\
    & $SO(3,1)_+\times SO(3,1)_+$ &$\begin{array}{l} f_{\alpha\,789} = -f_{\alpha\,129} = f_{\alpha\,183} = f_{\alpha\,723} = g_1\\ f_{\beta\,10,11,12} = -f_{\beta\,4,5,12} =g_2\\ f_{\beta 4,11,6} = f_{\beta\,10,5,6} = g_2\end{array}$   \\&&\\
    & $SO(2,1)_+\times SO(4,1)_+$ &$\begin{array}{l} f_{\alpha\,237} = f_{\alpha\,248} = f_{\alpha\,2,5,10} = f_{\alpha\,349}=g_1\\ 
    f_{\alpha \,3,5,11} = f_{\alpha\,4,5,12} = g_1 \\
    f_{\alpha\,789} = f_{\alpha\,7,10,11} = f_{\alpha\,8,10,12} =-g_1\\ f_{\alpha\,9,11,12} = -g_1, \,\,
    f_{\beta \,1,6,13} = g_2\end{array}$ 
\\ &&\\
    & $SO(2,1)_+\times SU(2,1)_+$ &$\begin{array}{l} f_{\alpha\,129} = f_{\alpha\,138} = f_{\alpha\,147} = f_{\alpha\,248} = -g_1\\ f_{\alpha\,237} = f_{\alpha\,349} = g_1\\ f_{\alpha\,1,2,10} = f_{\alpha\,3,4,10} = -\sqrt{3}g_1 \\ f_{\alpha 789} = 2g_1, \,\,f_{\beta\,5,6,11} = g_2\end{array}$ \\
    \hline
\end{tabular}
\caption{Embedding tensor components and gauge groups for the two types of gaugings that yield $dS_4$ vacua as given in \cite{dS4}. Some of the coupling constants have been rescaled compared to the original ones used in \cite{dS4}.}\label{dS4-sum}
\end{center}
\end{table}
\end{small}
\clearpage
\begin{table}[htb!]
\begin{center}
\begin{tabular}{|c|cl|}
\hline
& Gauge group & $\begin{array}{c} \text{Scalar potential $V(\phi, \chi$)} \\ \text{and $g_1/g_2$ scaling} \\ \text{for $dS_4$ vacuum at $\phi = \chi=0$}\end{array}$
\\\hline &&\\
\multirow{2}{*}{Type I} 
&
 $\begin{array}{l} SO(3)\times SO(3)\\\\
SO(3,1)\times SO(3,1)\\
\\ SO(3)\times SO(3,1)\\
\\
SO(3)\times SL(3, \mb R) \end{array}$ & $\begin{array}{c} 2 g_1 g_2 -\dfrac{1}{2}e^{-\phi}g_2^2 - \dfrac{1}{2}e^\p (g_1^2 + g_2^2\c^2) \\ g_1 = + g_2\,\,\,\,(dS_4)\\ g_1 = -g_2\,\,\,\,(AdS_4)
\end{array}$
\\ &&\\
\hline
    \multirow{18}{*}{$\begin{array}{c}\text{Type II}\end{array}$}
    &
    $SO(2,1)\times SO(2,1)$  & $\begin{array}{c}\dfrac{1}{2} e^{-\phi}\lf[g_2^2 + e^{2\phi}(g_1^2 + g_2^2\chi^2)\rr] \\ g_1 = \pm g_2\end{array}$
    \\&&\\
    & $ SO(2,1)\times SO(2,2)$& $\begin{array}{c}\dfrac{1}{2} e^{-\phi}\lf[ g_2^2  + e^{2\phi}\lf\{2g_1^2 + g_2^2 \chi^2\rr\}\rr]\\ g_1 = \pm\dfrac{1}{\sqrt{2}}g_2\end{array}$
    \\&&\\
    &$SO(2,1)\times SO(3,1)$& $\begin{array}{c} \dfrac{1}{2}e^{-\p}\lf[g_2^2 + e^{2\p}\lf(3g_1^2 + g_2^2\chi^2\rr) \rr]\\ g_1 = \pm \dfrac{1}{\sqrt{3}} g_2\end{array}$
  \\&&\\
&$SO(3,1)\times SO(3,1)$& $\begin{array}{c}\dfrac{3}{2}e^{-\p}\left(g_2^2 + e^{2\p}\left[g_1^2 + g_2^2\c^2\right]\right)\\ g_1 = \pm g_2\end{array}$
\\&&\\
    & $SO(2,1)\times SO(4,1) $
    &  $\begin{array}{c}\dfrac{1}{2}e^{-\p}\lf[g_2^2 + e^{2\p}\lf(6g_1^2 + g_2^2\chi^2\rr)\rr]\\ g_1 = \pm \dfrac{1}{\sqrt{6}} g_2 \end{array}$\\&&\\
    & $SO(2,1)\times SU(2,1)$ & $\begin{array}{c}\dfrac{1}{2}e^{-\p}\lf[g_2^2 + e^{2\p}\lf(12g_1^2 + g_2^2\chi^2\rr)\rr]\\\ g_1 = \pm \dfrac{1}{\sqrt{12}} g_2 \end{array}$\\ &&\\\hline
\end{tabular}
\caption{Scalar potentials constructed from the embedding tensor $f_{\a MNP}$ given in Table \ref{dS4-sum} using (\ref{Vs}) or (\ref{V-A}) for all gauge groups. For concreteness, we use $g_2$ for magnetic gauge factor $G_m$ and $g_1$ for electric factor $G_e$. Note that the magnetic gauge coupling always appears with $\chi^2$.}\label{dS4-sum-V}
\end{center}
\end{table}
\clearpage
\newpage
\section{$dS_2\times \Sigma_2$ solutions from 4D $N=4$ supergravity}\label{4D-dS2}
To obtain cosmological solutions interpolating between a $dS_2\times \S_2$ solution and a $dS_4$ solution in each of the gauge groups listed in Table \ref{dS4-sum-V} above, we need to turn on an Abelian $U(1)$ gauge field, together with the metric and supergravity scalars $\phi, \chi$. All other fields are truncated out. Specifically,
\beq
\phi^{m a} = 0, \qquad M^{MN} = M_{MN} = \mathbf 1_{6+n} \label{field-simp}
\eeq
Accordingly, the full Lagrangian (\ref{L-full}) reduces to the following general form
\beq e^{-1}\mc L &=& \frac{1}{2}R  - \frac{1}{4\,(\text{Im}\,\tau)^2}\partial_\m \tau \partial^\m \tau^* -\frac{1}{4}\text{Im}\,\tau\, \mc H^{M+}_{\m\n} \mc H^{M+\m\n}  
\non
&&\hspace{4mm} +\,\,\frac{1}{8}\,\text{Re}(\tau) \,\eta_{MN}\ep^{\m\n\rho\lambda}\mc H^{M+}_{\m\n} H_{\rho\lambda}^{N+} - V\,\,.\label{L-red-1}
\eeq
However, this Lagrangian (\ref{L-red-1}) will not be the final one that we will work with. 
In particular, for the $dS_2\times \S_2$ solutions that  are of interest to us, we will turn on the gauge field $U(1)_\text{diag} \subset U(1)_e\times U(1)_m \subset G_{e} \times G_\text{m}$. Since all our gaugings are dyonic, we will dualize the magnetic gauge factor of each gauging into an electric one using the procedure outlined in \cite{de-Wit-05-em}, so that the $SL(2,\mb R)$ frame under consideration is purely electric. This is necessary for us in order to use the Lagrangian (\ref{L-red-1}) in which only $SL(2,\mb R)$ electric field strengths are present. Effectively, this means that instead of $G_e\times G_m$ we will have $G_e\times \tilde G_e$ where $\tilde G_e$ is the dualized $G_m$. The dualization in this case is simply the following  $SL(2,\mb R)$ transformation acting on $\tau$ as \footnote{A general $SL(2,\mb R)$ transformation acting on $\tau$ has the form 
\[\tau \ra \tau' = \frac{ a\tau +b}{c\tau +d}, \qquad ad-bc = 1\]
so in the case of (\ref{sl2-duality}) $a = d = 0, \,\,b=1, \,c=-1$.}
\beq
\tau \ra \tau'=-\frac{1}{\tau}, \label{sl2-duality}
\eeq
so that the part of the Lagrangian (\ref{L-red-1}), involving the two gauge groups $G_e \times \tilde G_e$, takes the following specific form
\beq
e^{-1}\mc L_\text{gauge} &=& -\frac{1}{4} e^{-\p}\, F^{M}_{\m\n} F^{M\m\n}  +\,\,\frac{1}{8}\,\chi \,\eta_{MN}\ep^{\m\n\rho\lambda}F^{M}_{\m\n} F_{\rho\lambda}^{N}\non
&& -\frac{1}{4} \lf(\frac{e^{\p}}{1+\chi^2 e^{2\p}}\rr)\, \tilde F^{M}_{\m\n} \tilde F^{M\m\n}  -\,\,\frac{1}{8}\lf(\frac{\chi e^{2\phi}}{1+ \chi^2 e^{2\phi}} \rr)\,\eta_{MN}\ep^{\m\n\rho\lambda}\tilde F^{M}_{\m\n} \tilde F_{\rho\lambda}^{N}, \label{Lgauge}
\non
\eeq
with $F^M, \tilde F^M$ being the field strengths corresponding to $U(1)\subset G_e$ and $(1)\subset\tilde G_e$, respectively
\beq
F^{M} = 2\pd_{[\m}A^{M}_{\n]}, 
\qquad
\tilde F^{M} = 2\pd_{[\m}\tilde A^{M}_{\n]} 
\,\,.
\eeq
 Taking into account (\ref{Lgauge}), the Lagrangian (\ref{L-red-1}) becomes\footnote{This is the same as the Lagrangian given in \cite{Cvetic-Lu-Pope-99} where the case $SU(2)\times SU(2)$ gauge group was considered.}
\beq 
e^{-1}\mc L &=& \frac{1}{2}R  - \frac{1}{4\,(\text{Im}\,\tau)^2}\partial_\m \tau \partial^\m \tau^* + e^{-1}\mc L_\text{gauge} - V\,\,.\label{L-red-1b}
\eeq
\\
Next, we will truncate the axion $\chi$ in (\ref{L-red-1b}). This axion truncation is consistent as long as the following terms in (\ref{Lgauge}) which source $\chi$ vanish
\beq
\eta_{MN}\ep^{\m\n\rho\lambda}F^{M}_{\m\n} F_{\rho\lambda}^{N} = 0, \qquad \eta_{MN}\ep^{\m\n\rho\lambda}\tilde F^{M}_{\m\n} \tilde F_{\rho\lambda}^{N} = 0.
\eeq
This is the case because of the purely magnetic gauge ansatz that we will use, so $\chi$ can be safely truncated out.
\\\indent
Consequently, setting $\chi=0$ in (\ref{L-red-1b}) gives us the following Lagrangian that we will work with
\beq
e^{-1} \mc L = \frac{1}{2}R - \frac{1}{4}\pd_\m \phi \,\pd^\m \phi - \frac{1}{4}e^{-\p}F^{M}_{\m\n}F^{M\,\m\n}  - \frac{1}{4}e^\p \tilde F^{M}_{\,\m\n} \tilde F^{M\m\n} - V(\phi)\,\,. \label{L-fin}
\eeq
The explicit scalar potentials $V(\phi)$ for all gauge groups are given by those specified in Table \ref{dS4-sum-V} with $\chi=0$. 
\\\indent
Having established the Lagrangian (\ref{L-fin}), we now move on to specify the various ansatze for the $dS_2\times \S_2$ solutions. 
For the metric, the ansatz is
\beq
ds^2 = -dt^2 + e^{2f(t)}dr^2 + e^{2g(t)}\,d\Omega^2_2 \label{dS2-g}
\eeq
where $d\Omega^2_2$ is the line element for $S^2$ or $H^2$, 
\beq
d\Omega^2_2 = \begin{dcases}d\theta^2 + \sin^2\theta \,d\phi^2, &\S_2 = S^2\\ d\theta^2 + \sinh^2 \,d\phi^2, & \S_2 = H^2 \end{dcases}\,\,.
\eeq
The ansatz for the Abelian $U(1)_\text{diag} \subset U(1)_e\times U(1)_m$ gauge fields is
\beq
&&A^M_{\,\phi} = \tilde A^M_{\,\phi} = \begin{dcases} a \cos\theta, & \S_2 = S^2\\ a \cosh\theta, & \S_2 = H^2\end{dcases}\,\,. \label{dS2-A-ans}
\eeq
The exact gauge field ansatz with the specified values\footnote{$M$ assumes different values for $A$ and $\tilde A$.} for $M$ will be given in subsequent sections for each gauge group.\footnote{At this stage, although we only have $SL(2,\mb R)$ electric field strengths in the Lagrangian, we will continue to refer to the gauge ansatz as being either electric or magnetic when specifying $M$ later in each gauged theory. } For type I gaugings, we need to turn on the gauge fields corresponding to the $U(1)\times U(1)$ subgroup of the $SO(3)\times SO(3)$ compact subgroups  that lie entirely along R-symmetry directions, so $M=m$. For type II gaugings, the $U(1)\times U(1)$ have to be the subgroup of the compact parts that are embedded in the matter symmetry directions, so $M = a$. 
The corresponding gauge field strengths to (\ref{dS2-A-ans}) read
\beq
 F^{M}_{\theta\phi} =  \tilde F^{M}_{\theta\phi} = \begin{dcases} a \sin\theta, & \S_2 = S^2\\ a \sinh\theta, & \S_2 = H^2\end{dcases} \,\,.\label{dS2-F-ans}
\eeq
\\\indent 
The equations of motion for $dS_2\times \S_2$ solutions resulting from using the ansatze (\ref{dS2-g}, \ref{dS2-A-ans}) in the Lagrangian (\ref{L-fin}) are
\beq
0&=& \lambda\,e^{-2 g}-\frac{1}{2} a^2 e^{-4 g}\lf(e^{-\p} + e^\p\rr)+2 \ddot g+3 \dot g^2-V(\phi)+\frac{1}{4} \dot\phi^2,
\non
0&=& \frac{1}{2} a^2 e^{-4 g}\lf(e^{-\p} + e^\p \rr) +\ddot f+\dot f \dot g+\dot f^2+\ddot g+\dot g^2-V(\phi)+\frac{1}{4} \dot\phi^2,
\non
0&=& a^2 e^{-4 g}\lf(e^{-\p} - e^\p \rr)-\dot f \dot\phi-2 \dot g \dot\phi-2 V'(\phi)-\ddot \phi
\label{dS2eom}
\eeq
with 
\beq
\lambda = \begin{dcases}+1, &\S_2 = S^2\\
-1, & \S_2 = H^2 \end{dcases}\,\,.
\eeq
A $dS_2\times \S_2$ fixed-point solution of the equations (\ref{dS2eom}) is given by
\beq
\phi(t) = \phi_0,  \qquad g(t) = g_0, \qquad f(t) = f_0 t. 
\eeq
The full solution described by (\ref{dS2-g}) and $\phi = \phi(t)$ of (\ref{dS2eom}) is a cosmological solution interpolating between the above $dS_2\times \S_2$ fixed point at early times $t\ra -\infty$ and a $dS_4$ fixed point at late times $t\ra +\infty$. 
\\\indent
Before moving on, we note that real solutions can only be obtained if the product $a^2 g_1^2$ of the gauge flux $a$ and gauge coupling $g_1$ is negative. In particular, we will impose the following constraint
\beq
a\, g_1 = \pm \frac{i}{2}\,\,. \label{dS2twist}
\eeq
This situation resembles the case of cosmological solutions in those $dS$ supergravities, arising from dimensionally reducing the exotic $\star$-theories, with the wrong sign for the gauge field strengths \cite{Lu-03}, and is similar to the result obtained from the 5D analyses in 5D $N=4$ supergravity done in \cite{dS5-cosmo}. 
\newpage
\section{Type I $dS$ gauged theories}\label{4D-dS2-t1}
There are four gauge groups in the first type of $dS$ gaugings, namely $SO(3)\times SO(3)$, $SO(3,1)\times SO(3,1)$, $SO(4)\times SO(3)$, and  $SO(3)\times SL(3,\mb R)$.
These four gauged theories can give rise to both $dS_4$ and $AdS_4$ solutions.  The $SU(2)\times SU(2)$ common subgroup of the four gaugings are generated by $X_1, X_2, X_3$ and $X_4, X_5, X_6$, as can be seen from the embedding tensor components given in Table \ref{dS4-sum}.
 When the field content is truncated to only the metric and the scalars $\phi, \chi$ from the supergravity multiplet, all four gauged theories produce the same scalar potential 
\beq
V = 2 g_1 g_2 -\frac{1}{2}e^{-\phi}g_2^2 - \frac{1}{2}e^\p \lf(g_1^2  + g_2 \chi^2\rr).\label{V3131}
\eeq
that admits the following vacua at $\phi=\chi = 0$
\beq
\begin{array}{lll}
AdS_4: & g_1 = -g_2 & V_0 = -3g_1^2\\
dS_4: & g_1 = g_2 & V_0 = g_1^2.
\end{array} \label{t1-v}
\eeq
\subsection{$SO(3)\times SO(3)$}
\begin{figure}[!htb]
\centering 
  \begin{subfigure}[b]{0.45\textwidth}
    \includegraphics[width=\textwidth]{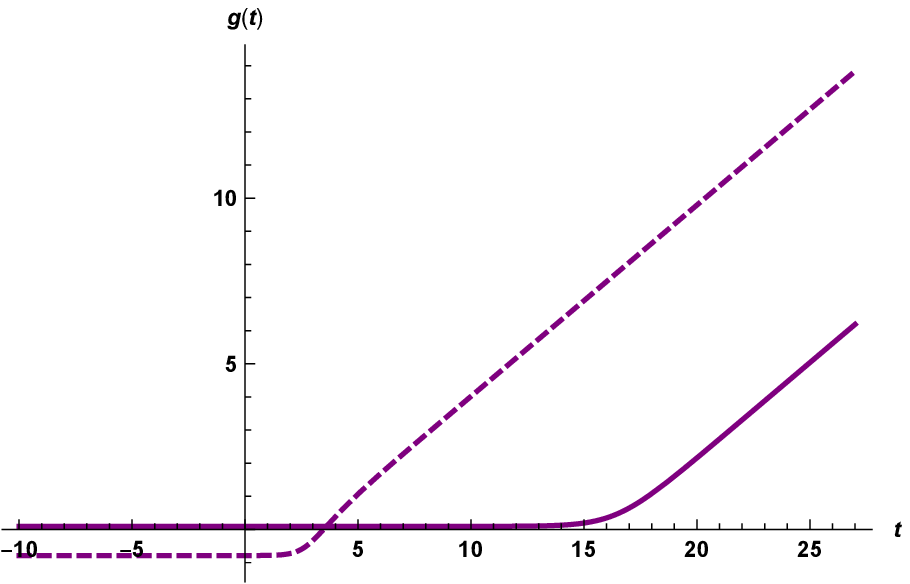}
\caption{Solution for $g$}  
  \end{subfigure}
   \begin{subfigure}[b]{0.45\textwidth}
    \includegraphics[width=\textwidth]{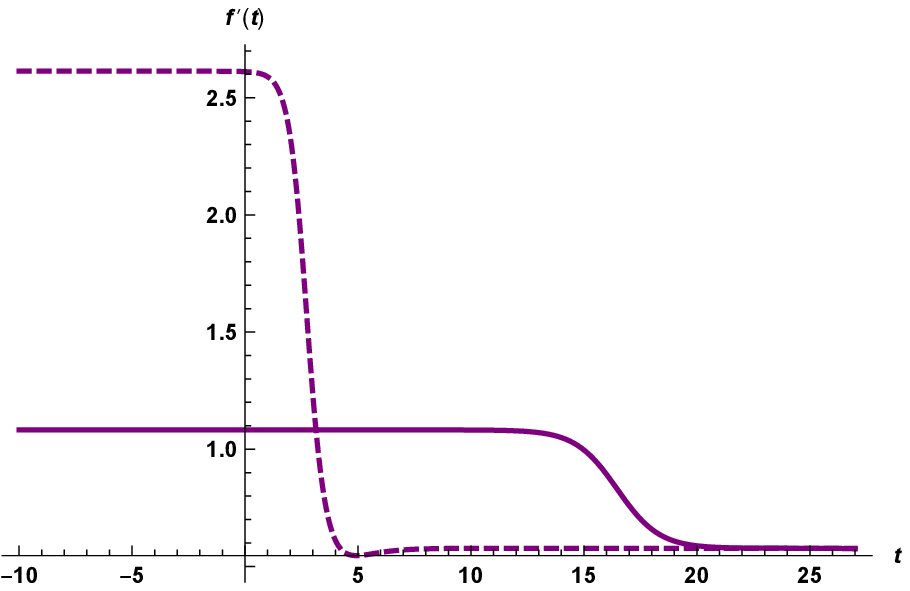}
\caption{Solution for $\dot f$}
\end{subfigure}      
    \caption{Cosmological solutions interpolating between the $dS_2\times \S_2$ (\ref{t1-so3131-dS2})fixed point (with solid line for $S^2$ and dashed line for $H^2$) at early times and the $dS_4$  (\ref{so31-t1-dS4})  solution at late times in the first type $dS$ gauging with gauge group $SO(3)\times SO(3)$ with $g_1 = 1$. }
    \label{fig:so31-t1-dS2}
    \end{figure}

This gauge group can be embedded entirely in the R-symmetry group $SO(6)$ without the need for any coupled matter. It is the simplest and only fully compact gauging for both $dS$ and $AdS$ solutions. 
To get $dS_4$ solution, we will set $g_1 = g_2$ in (\ref{V3131}) and also $\chi=0$ so that the scalar potential of the type I $dS$ gauged theory is
\beq
V =-\frac{1}{2} g_1^2 e^{-\phi } \left(1-4 e^{\phi }+e^{2 \phi }\right)
\label{so3131d}
\eeq
with the following $dS_4$ solution
\beq
\phi_0  = 0, \qquad f_0 = \frac{g_1}{\sqrt{3}}.
\label{so31-t1-dS4}
\eeq
As mentioned above, the $SO(3)\times SO(3)$ group is generated by $X_1, X_2, X_3$ and $X_4, X_5, X_6$. Therefore, the $U(1)\times U(1)$ gauge fields are given by (\ref{dS2-A-ans}) with $M=3$ for the electric part and $M=6$ for the magnetic part, corresponding to the generators $X_{3}$ and $X_{6}$, respectively. 
\\\indent
Eqs. (\ref{dS2eom}) together with the potential (\ref{so3131d}) yield the following $dS_2\times \S_2$ solutions
\beq
f_0&=& \,\sqrt{\frac{4 a^2 g_1^2+ \kappa\,\sqrt{1-4 a^2 g_1^2}-1}{2a^2}},
\non
g_0&=& -\frac{1}{2} \log \left(\frac{\kappa-\sqrt{1-4 a^2 g_1^2}}{2 a^2}\right),
\non
\phi_0&=& 0 \label{t1-so3131-dS20}
\eeq
with $\kappa = +1$ for $\S_2 = S^2$ and $\kappa = -1$ for $\S_2 = H^2$.
After imposing (\ref{dS2twist}), the solutions become
\beq
f_0 &=& \sqrt{2 (2-\kappa \sqrt{2})} \, g_1,\non
g_0 &=& -\frac{1}{2}\ln\lf[ 2(-\kappa + \sqrt{2}) \,g_1^2\rr]
\non
\phi_0 &=& 0, \label{t1-so3131-dS2}
\eeq
The cosmological solutions interpolating between the $dS_2\times \S_2$ fixed points (\ref{t1-so3131-dS2}) and the $dS_4$ solution (\ref{so31-t1-dS4}) are plotted in Fig.\ref{fig:so31-t1-dS2}.

\subsection{Other gauge groups}
In addition to the $SO(3)\times SO(3)$ theory above, type I $dS$ gauged theories include three more non-compact gauge groups $SO(3,1)\times SO(3,1)$, $SO(3)\times SO(3,1)$, and $SO(3)\times SL(3,\mb R)$ . These gauged theories cannot be realized with just pure $N=4$ supergravity. Instead, they require coupled vector multiplets in order to be implemented, since their non-compact directions are embedded entirely in the fundamental representation of the matter symmetry group $SO(n)\subset SO(6,n)$. However, because the $SO(3)\times SO(3)$ subgroup of these three theories are the same as the $SO(3)\times SO(3)$ gauge theory, the analyses for these three remaining gauge groups in the type I gauged theories  yield identical results to the $SO(3)\times SO(3)$ case studied above. As already mentioned, when truncated to just $\chi, \phi$, the scalar potentials as well as the gauge ansatze of all four theories are the same. Consequently, we will not repeat these analyses but note only that $dS_2\times \S_2$ fixed point solutions and cosmological solutions in these gauged theories are given by (\ref{t1-so3131-dS20}, \ref{t1-so3131-dS2}) and Fig. \ref{fig:so31-t1-dS2}, respectively. 
 
\section{Type II $dS$ gauged theories}\label{4D-dS2-t2}
For this type of gaugings, the $U(1)\times U(1)$ gauge fields are given by (\ref{dS2-A-ans})  with $M$ assuming the values along the matter symmetry  group $SO(n) \subset SO(6,n)$ in \ref{Mscalar} for both the electric and the magnetic parts. 
\subsection{$SO(2,1)\times SO(2,1)$}
\begin{figure}[!htb]
\centering 
  \begin{subfigure}[b]{0.45\textwidth}
    \includegraphics[width=\textwidth]{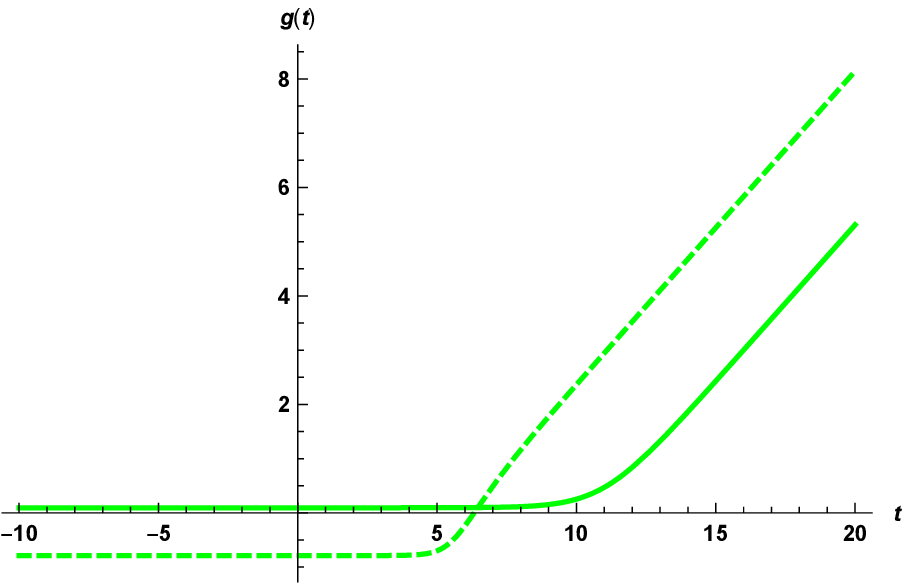}
\caption{Solution for $g$}  
  \end{subfigure}
   \begin{subfigure}[b]{0.45\textwidth}
    \includegraphics[width=\textwidth]{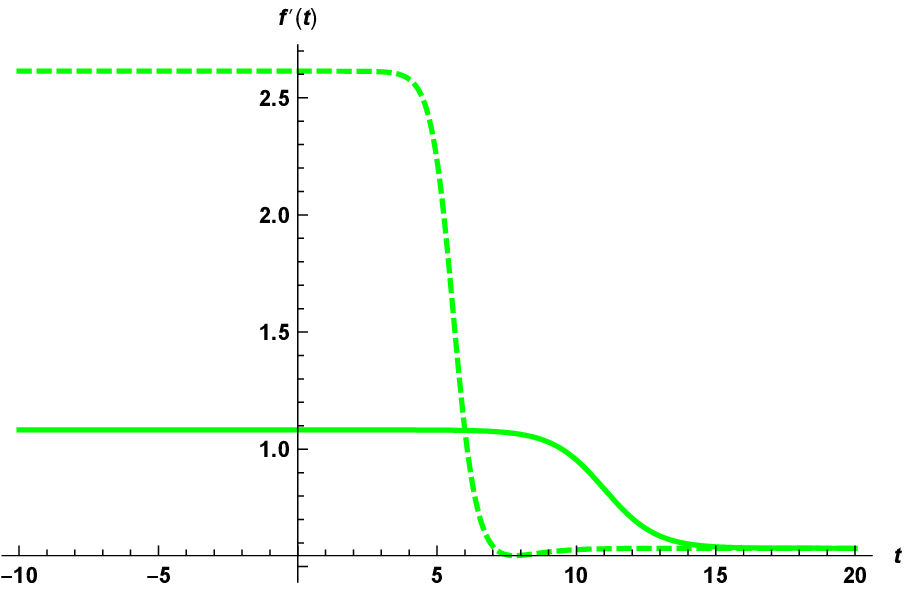}
\caption{Solution for $\dot f$}
\end{subfigure}      
    \caption{Cosmological solutions interpolating between the $dS_2\times \S_2$ (\ref{so21-dS2}) fixed point (with solid line for $S^2$ and dashed line for $H^2$) at early times and the $dS_4$ (\ref{so21-dS4}) solution at late times in the second type $dS$ gauging with gauge group $SO(2,1)\times SO(2,1)$ with $g_1 = 1$. }
    \label{fig:so21-t2-dS2}
    \end{figure}

From the embedding tensor given in Table \ref{dS4-sum}, the compact part $SO(2)\times SO(2)\subset SO(2,1)\times SO(2,1)$ is generated by $X_7$ and $X_{10}$. Accordingly, we can turn on the $U(1)\times U(1)$ gauge fields (\ref{dS2-A-ans}) with $M = 7$ and $M=10$ for the electric and magnetic parts, respectively. 
The scalar potential for this gauge group is 
\beq
V= \frac{1}{2}  \left(e^{\phi} g_1^2 + e^{-\phi} g_2^2\right)
\label{Vso2121}
\eeq
with a $dS_4$ vacuum at
\beq
\phi_0 =  0, \qquad f_0 = \frac{g_1}{\sqrt{3}}, \qquad g_1 = \pm g_2. \label{so21-dS4}
\eeq
The equations of motion (\ref{dS2eom}) together with the potential (\ref{Vso2121}) yield the following $dS_2\times \S_2$ fixed point solutions
\beq
f_0&=& \,\sqrt{\frac{4 a^2 g_1^2+ \kappa\,\sqrt{1-4 a^2 g_1^2}-1}{2a^2}},
\non
g_0&=& -\frac{1}{2} \log \left(\frac{\kappa-\sqrt{1-4 a^2 g_1^2}}{2 a^2}\right),
\non
\phi_0&=& 0\label{so21-dS2-0}
\eeq
which become
\beq
f_0 &=& \sqrt{2 (2-\kappa \sqrt{2})} \, g_1,\non
g_0 &=& -\frac{1}{2}\ln\lf[ 2(-\kappa + \sqrt{2}) \,g_1^2\rr]
\non
\phi_0 &=& 0 \label{so21-dS2}
\eeq
with $\kappa = 1$ for $\S_2 = S^2$ and $\kappa = -1$ for $\S_2 = H^2$, after imposing (\ref{dS2twist}).
\\\indent
The cosmological solutions interpolating between the $dS_2\times \S_2$ solutions (\ref{so21-dS2}) and the $dS_4$ solution (\ref{so21-dS4}) are plotted in Fig. \ref{fig:so21-t2-dS2}. 
   We note that the $dS_2\times \S_2$ fixed point solutions (\ref{so21-dS2-0}, \ref{so21-dS2}) and the cosmological solutions Fig. \ref{fig:so21-t2-dS2} are identical to those of the type I gauging, Eqs. (\ref{t1-so3131-dS20}, \ref{t1-so3131-dS2}) and Fig.(\ref{fig:so31-t1-dS2}).
\subsection{$SO(2,1)\times SO(2,2)$}
\begin{figure}[!htb]
\centering 
  \begin{subfigure}[b]{0.45\textwidth}
    \includegraphics[width=\textwidth]{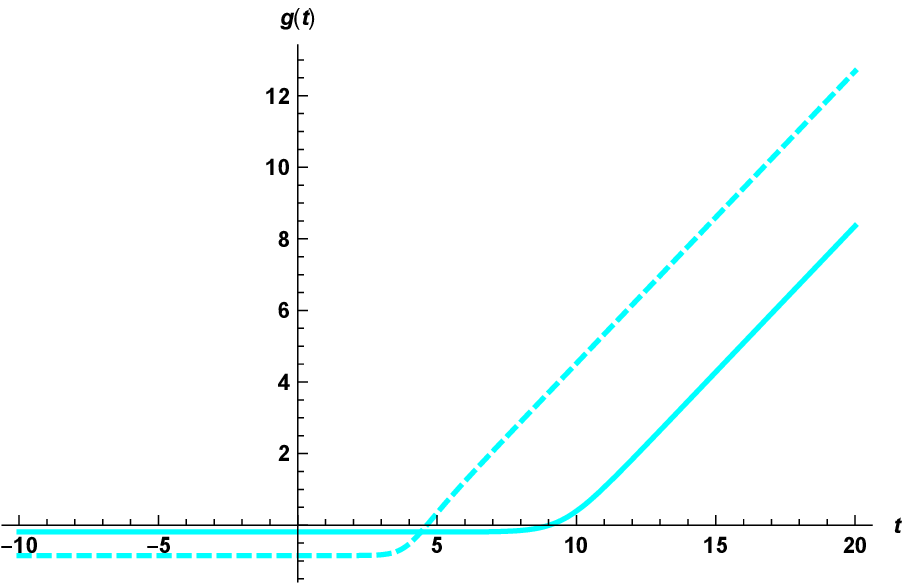}
\caption{Solution for $g$}  
  \end{subfigure}
   \begin{subfigure}[b]{0.45\textwidth}
    \includegraphics[width=\textwidth]{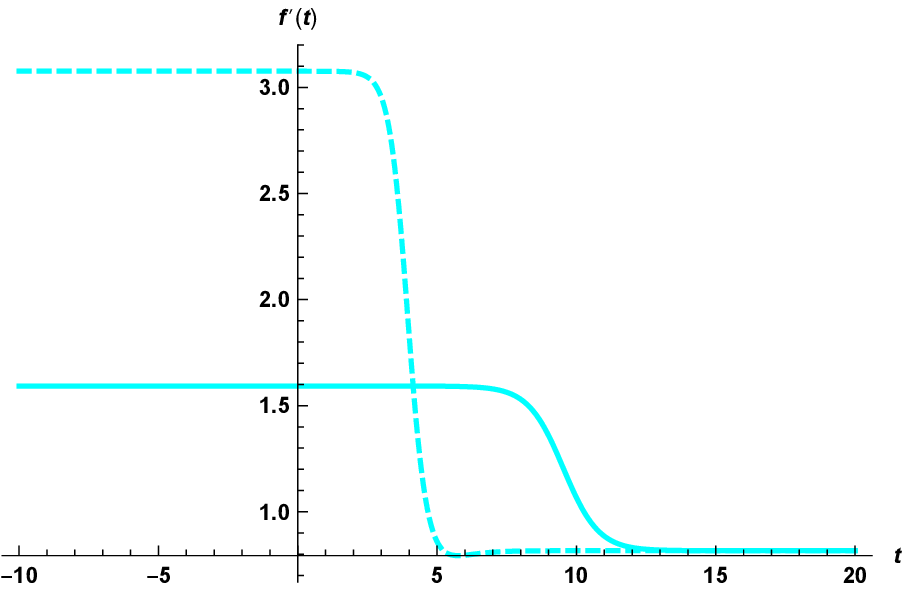}
\caption{Solution for $\dot f$}
\end{subfigure}      
    \caption{Cosmological solutions interpolating between the $dS_2\times \S_2$ (\ref{so22-dS2}) fixed point (with solid line for $S^2$ and dashed line for $H^2$) at early times and the $dS_4$ (\ref{so22-dS4}) solution at late times in the second type $dS$ gauging with gauge group $SO(2,1)\times SO(2,2)$ with $g_1 = 1$. }
    \label{fig:so22-t2-dS2}
    \end{figure}

From the embedding tensor given in Table \ref{dS4-sum}, the compact part $SO(2)\times SO(2)\times SO(2)\subset SO(2,1)\times SO(2,2)$ is generated by $X_9, X_8$ and $X_{7}$, respectively for each of the three $SO(2)$'s. Correspondingly, we can turn on the $U(1)\times U(1)$ gauge fields (\ref{dS2-A-ans}) with $M = 7$ or $M=8$ for the electric part and  $M=9$ for the magnetic part. 
The scalar potential of this theory is 
\beq
V 
=\frac{1}{2}  \left(e^{-\phi} g_2^2 + 2 g_1^2 \,e^{\phi}\right)\label{Vso22}
\eeq
with a $dS_4$ critical point at
\beq
\phi_0 =  0, \qquad f_0 = \sqrt{\frac{2}{3}}\,g_1, \qquad g_1 = \pm \frac{1}{\sqrt{2}} g_2. \label{so22-dS4}
\eeq
The equations (\ref{dS2eom}) with the potential $V$ (\ref{Vso22}) admit the following $dS_2\times \S_2$ fixed points
\beq
f_0&=& \,\sqrt{\frac{8 a^2 g_1^2+ \kappa\,\sqrt{1-8 a^2 g_1^2}-1}{2a^2}},
\non
g_0&=& -\frac{1}{2} \log \left(\frac{\kappa-\sqrt{1-8 a^2 g_1^2}}{2 a^2}\right),
\non
\phi_0&=& 0\label{so22-dS2-0}
\eeq
which become
\beq
f_0 &=& \sqrt{2 (3-\kappa \sqrt{3})} \, g_1,\non
g_0 &=& -\frac{1}{2}\ln\lf[ 2(-\kappa + \sqrt{3}) \,g_1^2\rr]
\non
\phi_0 &=& 0 \label{so22-dS2}
\eeq
with $\kappa = 1$ for $\S_2 = S^2$ and $\kappa = -1$ for $\S_2 = H^2$, after imposing (\ref{dS2twist}).
\\\indent
The cosmological solutions interpolating between the $dS_2\times \S_2$ solutions (\ref{so22-dS2}) and the $dS_4$ solution (\ref{so22-dS4}) are plotted in Fig. \ref{fig:so22-t2-dS2}. 

\subsection{$SO(3,1)\times SO(3,1)$}
\begin{figure}[!htb]
\centering 
  \begin{subfigure}[b]{0.45\textwidth}
    \includegraphics[width=\textwidth]{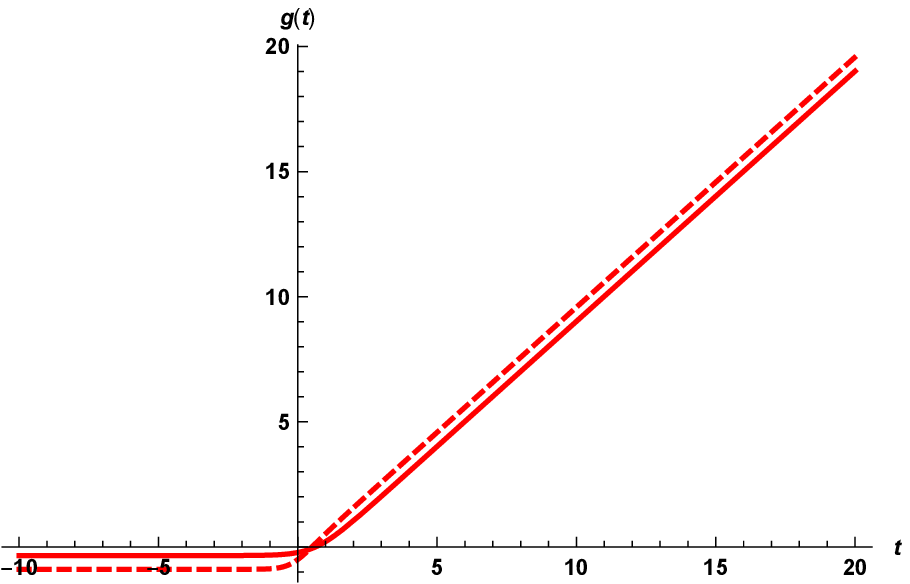}
\caption{Solution for $g$}  
  \end{subfigure}
   \begin{subfigure}[b]{0.45\textwidth}
    \includegraphics[width=\textwidth]{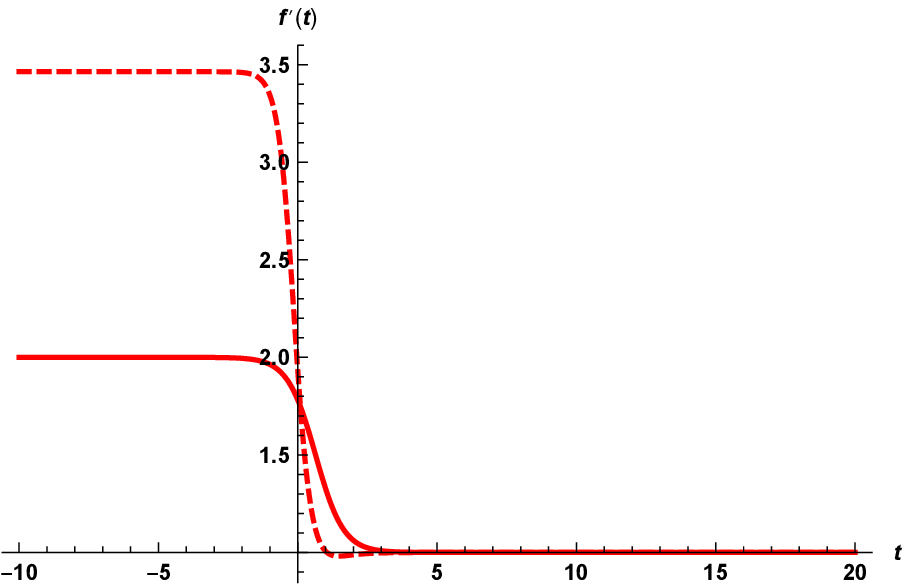}
\caption{Solution for $\dot f$}
\end{subfigure}      
    \caption{Cosmological solutions interpolating between the $dS_2\times \S_2$ (\ref{so31-dS2}) fixed point (with solid line for $S^2$ and dashed line for $H^2$) at early times and the $dS_4$ (\ref{so31-dS4}) solution at late times in the second type $dS$ gauging with gauge group $SO(3,1)\times SO(3,1)$ with $g_1 = 1$. }
    \label{fig:so3131-dS2-type2}
    \end{figure}

The compact part $SO(3)\times SO(3)\subset SO(3,1)\times SO(3,1)$ is generated by $X_7, X_8, X_9$ and $X_{10}, X_{11}, X_{12}$ (see Table \ref{dS4-sum}), so we can turn on the $U(1)\times U(1)$ gauge fields (\ref{dS2-A-ans}) with $M = 9$ for the electric part and $M=12$ for the magnetic part. 
The scalar potential for this gauge group is
\beq
V
=\frac{3}{2}\left(e^{ \phi} g_1^2+ g_2^2\, e^{-\phi} \right)
\label{Vso3131}
\eeq
with a $dS_4$ vacuum at
\beq
\phi_0 =  0, \qquad f_0 = g_1, \qquad g_1 = \pm g_2. \label{so31-dS4}
\eeq
The equations of motion (\ref{dS2eom}) together with the potential (\ref{Vso3131}) yield the following $dS_2\times \S_2$ fixed point solution
\beq
f_0&=& \sqrt{\frac{12 a^2 g_1^2+ \kappa\sqrt{1-12 a^2 g_1^2}-1}{2a^2}},
\non
g_0&=& -\frac{1}{2} \log \left(\frac{\kappa -\sqrt{1-12 a^2 g_1^2}}{2 a^2}\right),
\non
\phi_0&=& 0 
\label{so31-dS2-0}
\eeq
where $\kappa = 1$ for $\S_2 = S^2$ and $\kappa = -1$ for $\S_2 = H^2$.
 After imposing the condition (\ref{dS2twist}), (\ref{so31-dS2-0}) become
\beq
\begin{array}{llll}
\kappa = 1: &  f_0 = 2 g_1, \,\,\,& g_0 = -\log(\sqrt{2} g_1), \,\,\, &\phi_0 =0\\
\kappa = -1: &f_0 = 2\sqrt{3} g_1, \,\,\,& g_0 = -\log(\sqrt{6} g_1),\,\,\, & \phi_0 = 0.
\end{array}
\label{so31-dS2}
\eeq
The cosmological solutions interpolating between (\ref{so31-dS2}) at early times and (\ref{so31-dS4}) at late times are numerically solved for and plotted in Fig.\ref{fig:so3131-dS2-type2}. 
\subsection{$SO(2,1)\times SO(3,1)$}
The compact part $SO(2)\times SO(3)\subset SO(2,1)\times SO(3,1)$ is generated by $X_{10}$ and $X_{7}, X_8, X_9$, respectively (see Table \ref{dS4-sum}). Accordingly, we can turn on the $U(1)\times U(1)$ gauge fields (\ref{dS2-A-ans}) with $M = 9$ for the electric part and $M=10$ for the magnetic part. 
The scalar potential of this theory is
\beq
V = \frac{1}{2}\left(3\,e^{ \phi} \, g_1^2+ g_2^2\,e^{-\phi}\right) \label{so3122-V}
\eeq
with the following $dS_4$ vacuum
\beq
\phi_0 = 0, \qquad f_0 =  g_1, \qquad g_1 = \pm \frac{1}{\sqrt{3}}g_2.
\eeq
which has the same $f_0$ as (\ref{so31-dS4}).
The equations (\ref{dS2eom}) with the potential (\ref{so3122-V}) yield the same $dS_2\times \S_2$ fixed-point solutions as (\ref{so31-dS2-0}) of the $SO(3,1)\times SO(3,1)$ theory, which become (\ref{so31-dS2}) after imposing (\ref{dS2twist}). The cosmological solutions from this theory, interpolating between (\ref{so31-dS2}) and (\ref{so3122-V}), are the same as and given by Fig.\ref{fig:so3131-dS2-type2}.
\subsection{$SO(2,1)\times SO(4,1)$}
\begin{figure}[!htb]
\centering 
  \begin{subfigure}[b]{0.45\textwidth}
    \includegraphics[width=\textwidth]{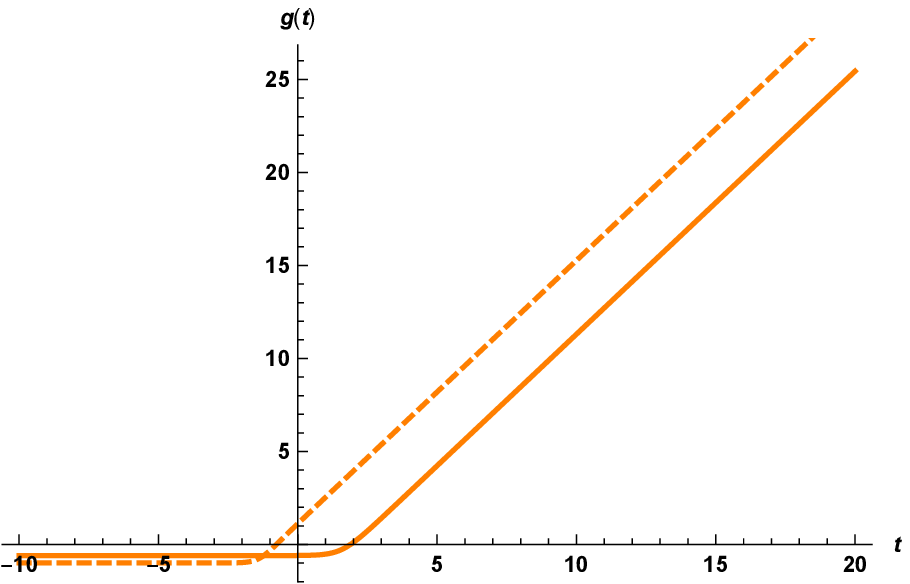}
\caption{Solution for $g$}  
  \end{subfigure}
   \begin{subfigure}[b]{0.45\textwidth}
    \includegraphics[width=\textwidth]{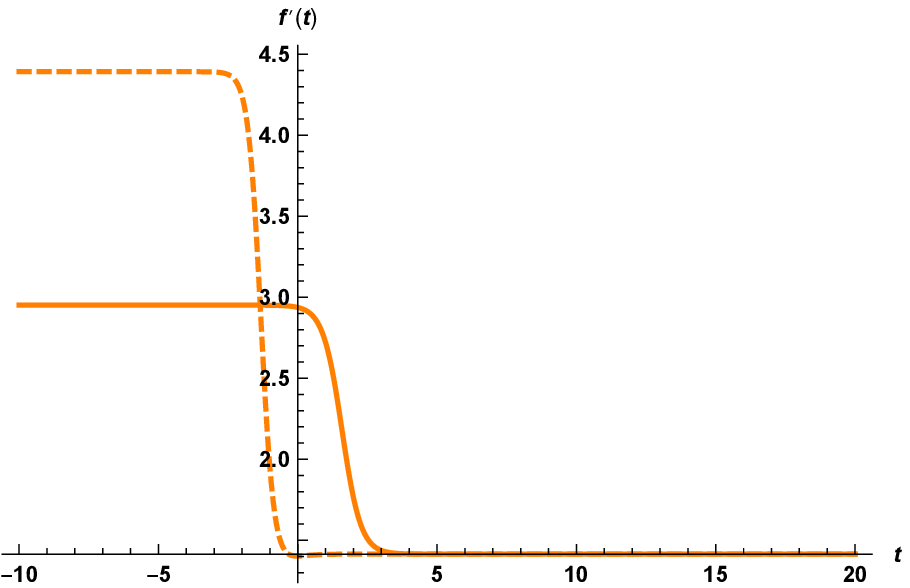}
\caption{Solution for $\dot f$}
\end{subfigure}      
    \caption{Cosmological solutions interpolating between the $dS_2\times \S_2$ (\ref{so41-dS2}) fixed point (with solid line for $S^2$ and dashed line for $H^2$) at early times and the $dS_4$ (\ref{so41-dS4}) solution at late times in the second type $dS$ gauging with gauge group $SO(2,1)\times SO(4,1)$ with $g_1 = 1$. }
    \label{fig:so41-t2-dS2}
    \end{figure}

The compact part $SO(2)\times SO(4)\subset SO(2,1)\times SO(4,1)$ is generated by $X_{13}$ and $X_{7}, \ldots, X_{12}$, respectively (see Table \ref{dS4-sum}). Equivalently, the $SO(4)$ factor can be written as $SO(4)\cong SO(3)_+\times SO(3)_-$ with the corresponding generators
\beq
SO(3)_\pm: \qquad X_7\pm X_{12}, \qquad X_8 \mp X_{11}, \qquad X_9\pm X_{10}\,\,. \label{so433}
\eeq
The $U(1)\times U(1)$ gauge fields in this case are given by (\ref{dS2-A-ans}) with $M=13$ for the magnetic part corresponding to $SO(2)\subset SO(2,1)$, and
\beq
A^{7}_\phi = A^{12}_\phi = \frac{a}{\sqrt{2}}\begin{dcases}\cos\theta & \S_2 = S^2\\ \cosh\theta & \S_2 = H^2 \end{dcases}
\eeq
for the electric part, corresponding to the Cartan of $SO(3)_+$ factor in (\ref{so433}). 
\\\indent
The scalar potential for this gauge group is
\beq
V
=\frac{1}{2} \left(6\,e^{ \phi} g_1^2+g_2^2\,e^{-\phi} \right)
\label{Vso2141}
\eeq
with a $dS_4$ vacuum at
\beq
\phi_0 =  0, \qquad f_0 = \sqrt{2}\,g_1, \qquad g_1 = \pm\frac{1}{\sqrt{6}}g_2. \label{so41-dS4}
\eeq
The equations of motion (\ref{dS2eom}) together with the potential (\ref{Vso2141}) yield the following $dS_2\times \S_2$ fixed point solutions
\beq
f_0&=& \sqrt{\frac{24 a^2 g_1^2+\kappa\sqrt{1-24 a^2 g_1^2}-1}{2a^2}},
\non
g_0&=& -\frac{1}{2} \log \left(\frac{\kappa -\sqrt{1-24 a^2 g_1^2}-1}{2 a^2}\right),
\non
\phi_0&=& 0
\label{so41-dS2-0}
\eeq
which, after imposing (\ref{dS2twist}), become
\beq
f_0&=& \sqrt{2} \sqrt{\left(7-\kappa\sqrt{7}\right) g_1^2}
\non
g_0&=& -\frac{1}{2} \log \left(2 \left(\sqrt{7}-\kappa\right) g_1^2\right),
\non
\phi_0&=& 0
\label{so41-dS2}
\eeq
where $\kappa = +1$ for $\S_2 = S^2$ and $-1$ for $\S_2 = H^2$.
The cosmological solutions connecting (\ref{so41-dS2}) at early times to (\ref{so41-dS4}) at late times are numerically solved for and plotted in Fig. \ref{fig:so41-t2-dS2}.

\subsection{$SO(2,1)\times SU(2,1)$}
\begin{figure}[!htb]
\centering 
  \begin{subfigure}[b]{0.45\textwidth}
    \includegraphics[width=\textwidth]{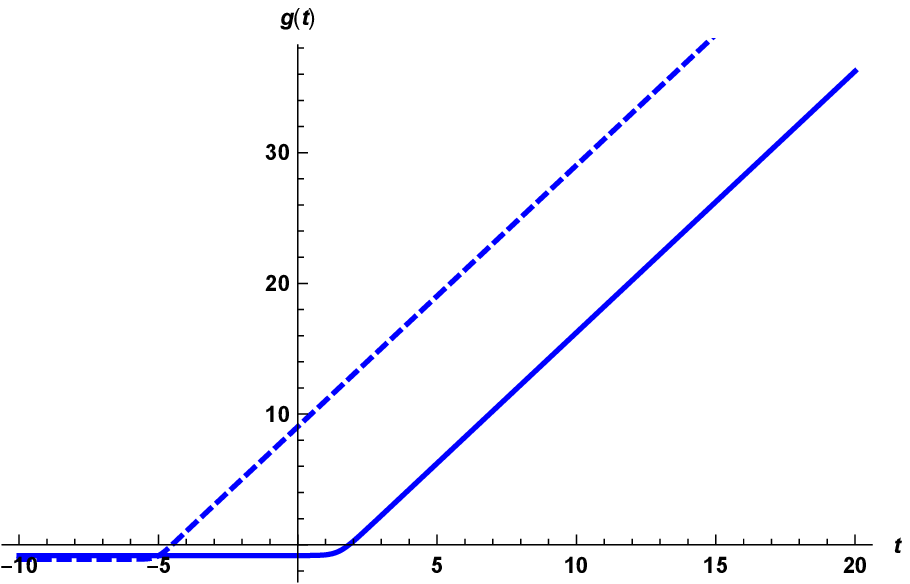}
\caption{Solution for $g$}  
  \end{subfigure}
   \begin{subfigure}[b]{0.45\textwidth}
    \includegraphics[width=\textwidth]{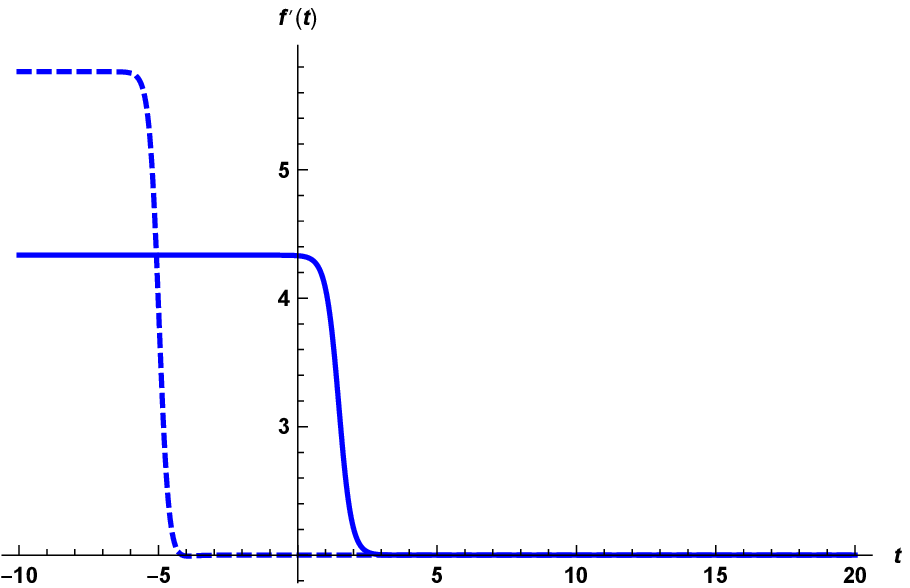}
\caption{Solution for $\dot f$}
\end{subfigure}      
    \caption{Cosmological solutions interpolating between the $dS_2\times \S_2$ (\ref{su21-dS2}) fixed point (with solid line for $S^2$ and dashed line for $H^2$) at early times and the $dS_4$ (\ref{su21-dS4}) solution at late times in the second type $dS$ gauging with gauge group $SO(2,1)\times SU(2,1)$ with $g_1 = 1$. }
    \label{fig:su21-t2-dS2}
    \end{figure}

The compact part $SO(2)\times SU(2)\times U(1)\subset SO(2,1)\times SU(2,1)$ is generated by $X_{11}$, $X_{7}, X_8, X_9$, and $X_{10}$, respectively (see Table \ref{dS4-sum}). Hence, we can turn on the $U(1)\times U(1)$ gauge fields (\ref{dS2-A-ans}) with $M = 11$ for the magnetic part and $M=9$ or $M=10$ for the electric part.
The scalar potential for this gauge group is
\beq
V
=\frac{1}{2} \left(12 g_1^2 \,e^{ \phi} +g_2^2\,e^{-\phi} \right)
\label{Vsu2121}
\eeq
with a $dS_4$ vacuum at
\beq
\phi_0 =  0, \qquad f_0 = 2g_1, \qquad g_1 = \pm\frac{1}{\sqrt{12}}g_2. \label{su21-dS4}
\eeq
The equations of motion (\ref{dS2eom}) together with the potential (\ref{Vsu2121}) yield the following $dS_2\times \S_2$ fixed point solutions
\beq
f_0&=& \sqrt{\frac{48 a^2 g_1^2+\kappa\sqrt{1-48 a^2 g_1^2}-1}{2a^2}},
\non
g_0&=& -\frac{1}{2} \log \left(\frac{\kappa-\sqrt{1-48 a^2 g_1^2}}{2 a^2}\right),
\non
\phi_0&=& 0
\label{su21-dS20}
\eeq
which, after imposing (\ref{dS2twist}), become
\beq
f_0&=& \sqrt{2} \sqrt{\left(13-\kappa\sqrt{13}\right) g_1^2},
\non
g_0&=& -\frac{1}{2} \log \left(2 \left(\sqrt{13}-\kappa\right) g_1^2\right),
\non
\phi_0&=& 0
\label{su21-dS2}
\eeq
where $\kappa = +1$ for $\S_2 = S^2$ and $-1$ for $\S_2 = H^2$.
The cosmological solutions connecting (\ref{su21-dS2}) at early times to (\ref{su21-dS4}) at late times are numerically solved for and plotted in Fig. \ref{fig:su21-t2-dS2}.
\subsection{Summary of all solutions}
\begin{figure}[!htb]
\centering 
   \begin{subfigure}[b]{0.6\textwidth}
    \includegraphics[width=\textwidth]{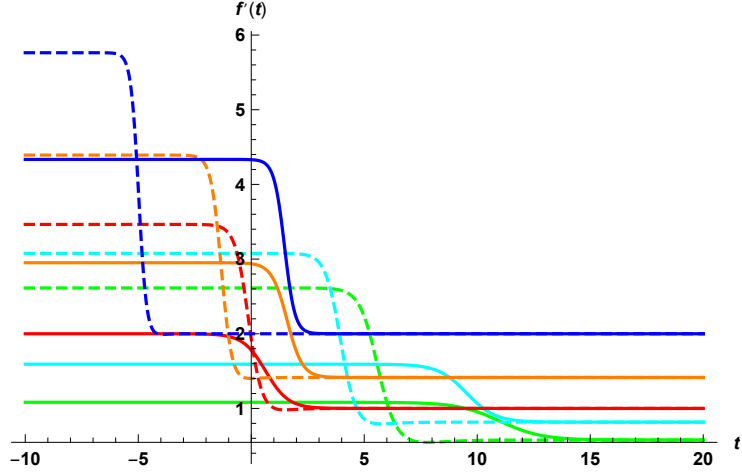}
\caption{Solution for $\dot f$}
\end{subfigure} 
      \begin{subfigure}[b]{0.85\textwidth}
    \includegraphics[width=\textwidth]{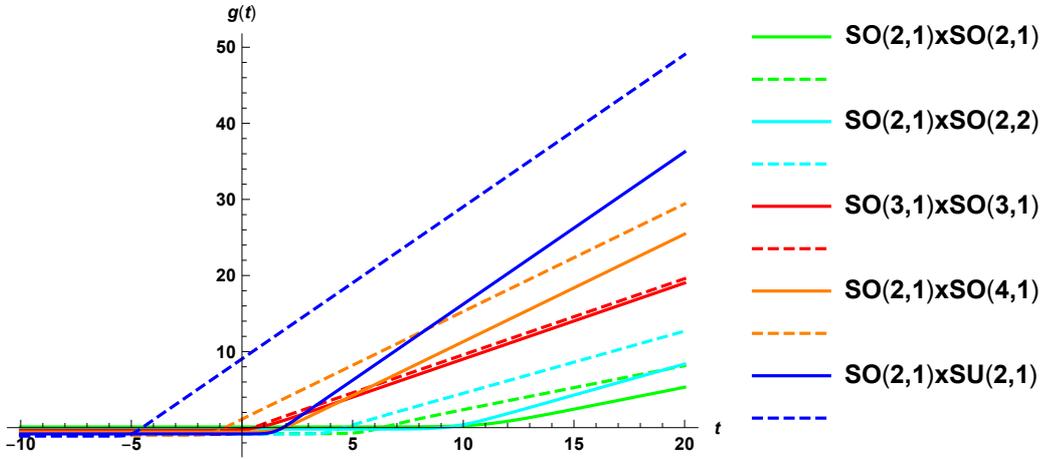}
\caption{Solution for $g$}
    \end{subfigure}
    \caption{All cosmological solutions from $dS_2\times \S_2$ at the infinite past to $dS_4$ in the infinite future  with $g_1 = 1$ from the type II $dS$ gauged theories. Solid lines represent $\S_2 = S^2$ solutions and dashed lines represent $\S_2 = H^2$ solutions. Note that the solutions of the $SO(2,1)\times SO(3,1)$ theory are given by those of $SO(3,1)\times SO(3,1)$ theory.}
    \label{fig:all-4DdS2-cf}
    \end{figure}

In this section, we summarize all solutions of the type II $dS$ gauged theories. For all gauge groups, excluding $SO(3,1)\times SO(3,1)$, the fixed-point solutions can be rewritten in a common form using the $g_1/g_2$ ratios given in Table \ref{dS4-sum-V}. We list all the rewritten solutions in Table \ref{table:dS2-all}. Furthermore, we collect all cosmological solutions given Figs. \ref{fig:so21-t2-dS2}, \ref{fig:so22-t2-dS2},  \ref{fig:so3131-dS2-type2}, \ref{fig:so41-t2-dS2}, \ref{fig:su21-t2-dS2} in a single plot Fig. \ref{fig:all-4DdS2-cf}.
\begin{table}[!htb]
\centering
\begin{tabular}{|c|c|}
\hline &\\
 Type II gauge group & $dS_2\times \S_2$ solutions\\ &\\\hline &\\
$\begin{array}{l} SO(2,1)\times SO(2,1)\\ \\ SO(2,1)\times SO(2,2)\\\\ SO(2,1)\times SO(3,1)\\\\
SO(2,1)\times SO(4,1)\\ \\SO(2,1)\times SU(2,1)\end{array}$ &
 $\begin{array}{l} f_0= \,\sqrt{\dfrac{4 a^2 g_2^2+ \kappa\,\sqrt{1-4 a^2 g_2^2}-1}{2a^2}},
\\
g_0= -\dfrac{1}{2} \log \left(\dfrac{\kappa-\sqrt{1-4 a^2 g_2^2}}{2 a^2}\right),\\
\end{array}$
\\&\\\hline &\\
$SO(3,1)\times SO(3,1)$ & $\begin{array}{l} f_0= \,\sqrt{\dfrac{12 a^2 g_1^2+ \kappa\,\sqrt{1-12 a^2 g_1^2}-1}{2a^2}},
\\
g_0= -\dfrac{1}{2} \log \left(\dfrac{\kappa-\sqrt{1-12 a^2 g_1^2}}{2 a^2}\right),
\end{array}$ \\ &
\\\hline
\end{tabular}
\caption{Summary of all fixed-point $dS_2\times \S_2$ solutions in type II $dS$ gauged theories with the six gauge groups. All solutions have $\phi_0 = 0$, and $\kappa = +1$ for $\S_2 = S^2$ and $\kappa = -1$ for $\S_2 = H^2$.}\label{table:dS2-all}
\end{table}
\newpage
\clearpage
\section{First-order systems}\label{BPS}
In this section, we will derive the first-order equations, containing
some relevant pseudo-superpotential $W$ \cite{ST06}, \cite{ST0610}, \cite{VR-07}, that can solve the second order equations of motion \ref{dS2eom}. Next, we check whether these first-order equations admit the $dS_4$ vacua and their associated cosmological solutions found in sections \ref{4D-dS2-t1} and \ref{4D-dS2-t2}. 
Although first-order equations are most often linked to the supersymmetric $AdS$ case in which  domain walls or holographic RG flow solutions can arise as solutions of some first-order BPS equations obtained by setting to zero the supersymmetry transformations of the fermionic fields, it must be noted that first-order equations can and do arise completely independently from supersymmetric systems in the so-called fake supersymmetric case. In this scenario, the Hamilton-Jacobi (HJ) formalism has been shown to produce first-order equations containing some fake superpotential obtained from the factorization of the HJ characteristic function \cite{VR-16}. More details on the relation between the HJ formalism and fake supersymmetry can be found in \cite{VR-12}, \cite{VR-16}. 
\\\indent
As mentioned in section \ref{intro}, the fact that $dS_4$ vacua of 4D $N=4$ supergravity are unstable is a strong indication of the lack of pseudosuperymmetry that ensures the existence of relevant pseudosuperpotentials and corresponding first-order equations. It is then the objective of this section to characterize the extent to which there fail to exist  relevant pseudosuperpotentials for the type I and type II $dS$ theories. We will proceed as follows. 
For the purpose of deriving the first-order equations in the $dS$ gauged theories, our analysis will be based on the $AdS$ case since their equations of motion are almost identical up to various signs. The form of the first-order equations for the $dS$ case will be inferred from that of the BPS equations in the $AdS$ case.  
To derive the field equations for either $AdS_4$ and $dS_4$ solutions, we will work with the minimally required field content consisting only of the metric and dilaton $\phi$. The axion $\chi$ can be truncated out since it vanishes in either $AdS_4$ or $dS_4$ vacuum. 
The Lagrangian reads
\beq
e^{-1}\mc L &=& \frac{1}{2}R  - \frac{1}{4}\partial_\m \p \,\partial^\m \p - V(\phi). \label{L-red}
\eeq
In the cases of the $AdS_2\times \S_2$ and $dS_2\times \S_2$, we will use (\ref{L-fin}).
\\\indent

\subsection{$AdS$ case}
\subsubsection{First-order equations for $AdS_4$}
The ansatz for $AdS_4$ metric is
\beq
ds^2 =  dr^2 +e^{f(r)} \lf[-dt^2 +dx^2 + dy^2\rr] \,.\label{AdS4}
\eeq
The field equations resulting from using (\ref{AdS4}) in (\ref{L-red}) are
\beq
0&=& 2 f''+3 f'^2+\frac{1}{4} \phi'^2+ V,
\non
0&=&\frac{1}{2}\phi''+\frac{3}{2} f' \phi'- \lf(\frac{\partial V}{\partial \phi}\rr),
\label{AdS4eom}
\eeq
The equations (\ref{AdS4eom}) are solved by the following set of first-order equations
\beq
f'&=& \frac{\sqrt{2}}{3} \,\, W,
\non
\phi'&=& -\frac{4\sqrt{2}}{3}\,\,\lf(\frac{\partial W}{\partial \phi}\rr),
 \label{AdS4bps}
\eeq
subject to the following condition on the scalar potential and the superpotential
\beq
V&=& \frac{8}{9}\lf(\frac{\partial W}{\partial \phi}\rr) ^2-\frac{2}{3} W^2\,\,.
\label{wv-A}
\eeq
Recall that there are four gaugings which admit fully supersymmetric $AdS_4$ solutions. These are the exact ones given in the type I $dS$ gaugings (see Table \ref{dS4-sum}). With the vector multiplet scalars truncated out, the scalar potentials from these four gaugings are identical and can be obtained from the ones in Table \ref{dS4-sum-V} by setting $\chi=0$.
\beq
V = 2 g_1 g_2 - \frac{1}{2}e^{-\p} g_2^2 - \frac{1}{2}e^\p  g_1^2 \label{V-2a}
\eeq
To obtain $AdS_4$ solutions we need to set $g_1 = -g_2$ so that the potential (\ref{V-2a}) becomes
\beq
V = -\frac{1}{2}g_1^2 e^{-\p}\lf( 1+4 e^\p + e^{2\p}  \rr) \label{V-a}
\eeq
There exists the following superpotential $W$
\beq
W =\frac{3g_1}{2\sqrt{2}} \lf(e^{-\p/2} + e^{\p/2}\rr) \label{w-A-a}
\eeq
such that the potential (\ref{V-a}) can be written in terms of $W$ as in the required relation (\ref{wv-A}). Note that the equations (\ref{AdS4bps}) can be derived by setting to zero the supersymmetry transformations of the gravitino and dilatino fields, $\d\psi_{\m\,i} = 0, \d\chi^i = 0$, as were done in \cite{KU-AdS4}\footnote{In \cite{KU-AdS4}, the derivation of the BPS equations, from $\d\psi_{\m\,i} = 0, \d\chi^i = 0$, involves the axion $\chi$. As such, the superpotential and first-order equations are more general. Another difference between \cite{KU-AdS4}  and this work is a factor of 2 in $V$ and the first-order equations given in \cite{KU-AdS4}.}. Before moving on, we remark that the BPS equations (\ref{AdS4bps}) with the superpotential $W$ (\ref{w-A-a}) and the equations of motion (\ref{AdS4eom}) with the scalar potential $V$ (\ref{V-a}) admit the same $AdS_4$ vacuum
\beq
f_0 = g_1, \qquad \phi_0 = 0, \label{AdS4}
\eeq
as should be the case. 

\subsubsection{First-order equations for $AdS_2\times \S_2$}
 The metric ansatz for $AdS_2\times \S_2$ is
\beq
ds^2 =  -e^{f(r)}\,dt^2+ dr^2 +e^{g(r)}\,d\Omega^2_2,\label{AdS2}
\eeq
with $d\Omega_2^2$ being the line element for $\S_2 =S^2, H^2$.
The gauge field ansatz is given by (\ref{dS2-A-ans}), with $A^M_{e(m)\,\phi}$ being the $U(1)\times U(1)$ gauge fields in the R-symmetry directions where $M=3$ for $\a = e$ and $M=6$  for $\a=m$. This is exactly the same as the gauge ansatz for the type I $dS$ gauged theories.  
The equations of motion for $AdS_2\times \S_2$ resulting from using the ansatze (\ref{AdS2}, \ref{dS2-A-ans}) in (\ref{L-fin}) are
\beq
0&=& -\lambda e^{-2 g}+\frac{1}{2} a^2 e^{-4 g}\lf(e^{-\phi}+ e^{\phi} \rr)+2 g''+3 g'^2+V(\phi)+\frac{1}{4} \phi'^2,
\non
0&=& -\frac{1}{2}a^2  e^{-4 g}\left(  e^{-\phi}+ e^{\phi}\rr)+ f''+ f' g'+ f'^2+ g''+g'^2+ V(\phi)+\frac{1}{4} \phi'^2,
\non
0&=& -a^2 e^{-4 g}\left( e^{-\phi}- e^{\phi}\rr) - f' \phi' - 2 g' \phi'+ 2 V'(\phi)-\phi'' \label{AdS2eom}
\eeq
with $\lambda = +1$ for $\S_2 = S^2$ and $\lambda = -1$ for $\S_2 = H^2$. It is instructive to compare the field equations (\ref{AdS2eom}) in this case  to those from the $dS$ case (\ref{dS2eom}). The two sets are almost identical except for the opposite signs for the non-derivative terms ($V, \lambda e^{-2g}$, and gauge field strength terms). 
The equations of motion for the $AdS_2\times S_2$ (\ref{AdS2eom}) case are solved by the following set of first-order system
\beq
f'&=& \frac{\sqrt{2}}{3} \, W\,\left(\frac{\lambda a e^{-2 g}}{g_1}+1\right),
\non
g'&=& \frac{\sqrt{2}}{3} \, W\, \left(1-\frac{\lambda a e^{-2 g}}{g_1}\right),\non
\phi'&=& -\frac{4\sqrt{2}}{3} \, \left(1-\frac{\lambda a e^{-2 g}}{g_1}\right) \,\frac{\partial W}{\partial \phi}, \label{AdS2bps}
\eeq
 subject to the  condition (\ref{wv-A}) on the scalar potential and the superpotential, and the following condition on the gauge flux $a$ and gauge coupling $g_1$
\beq
a g_1 = -\frac{1}{2}.\label{AdS2twist}
\eeq
For the scalar potential $V$ given by (\ref{V-a}), the relation (\ref{wv-A}) was shown to be satisfied with the superpotential given by (\ref{w-A-a}).
The BPS equations (\ref{AdS2bps}) subject to (\ref{AdS2twist}) with $W$ given by (\ref{w-A-a}) admit the same $AdS_2\times \S_2$ solutions as the equations of motion (\ref{AdS2eom}) with the potential (\ref{V-a}). However, not all solutions are real and thus physically acceptable.  In particular, only in the case $\lambda = -1$, there exists the following real  $AdS_2\times H_2$ solution to both (\ref{AdS2bps}) and (\ref{AdS2eom})
\beq
f_0 = 2 g_1, \qquad g_0 = -\frac{1}{2}\log\lf[ 2 g_1^2\rr], \qquad \phi_0 = 0 \label{AdS2-H2}
\eeq
The domain wall solution interpolating between the $AdS_2\times H_2$ solution (\ref{AdS2-H2}) and the $AdS_4$ solution (\ref{AdS4}) can be obtained by either solving (\ref{AdS2eom}) or (\ref{AdS2bps}) numerically. 
Finally, we note that the first-order equations (\ref{AdS2bps}) with $W$ given by (\ref{w-A-a}) and the twist condition (\ref{AdS2twist}) are essentially identical to the BPS equations that are given in \cite{Bobev-universal} for the 4D $N=4$ $AdS_2\times \S_2$ holographic RG flow solutions. 
\subsection{$dS$ case}
\subsubsection{First-order equations for $dS_4$}
The metric ansatz for $dS_4$ is
\beq
ds^2 = -dt^2 + e^{f(t)}\,(dx^2 + dy^2 + dz^2)\,.
\eeq
Using this ansatz in (\ref{L-red}) gives the following set of equations of motion
\beq
0 &=&2 \ddot f+3 \dot f^2+\frac{1}{4} \dot\phi^2 - V,
\non
0&=& \frac{1}{2}\ddot \phi+ \frac{3}{2} \dot f \dot\phi + \lf(\frac{\partial V}{\partial \phi}\rr)
\,. \label{dS4eom}
\eeq 
These equations are almost identical to the ones for $AdS_4$  (\ref{AdS4eom}), save for the opposite signs in front of the  terms involving the scalar potential and its derivative.
Eqs. (\ref{dS4eom}) can be solved by the same set of first-order equations (\ref{AdS4bps}) as in the $AdS_4$ case 
\beq
\dot f&=& \frac{\sqrt{2}}{3} \,\, W,
\non
\dot \phi&=& -\frac{4\sqrt{2}}{3}\,\,\lf(\frac{\partial W}{\partial \phi}\rr),
 \label{dS4bps}
\eeq
but with the relation (\ref{wv-A}) replaced by 
\beq
V&=& -\lf[\frac{8}{9}\lf(\frac{\partial W}{\partial \phi}\rr)^2-\frac{2}{3} W^2\rr],\,\, \label{vw-d}
\eeq
where $V$ has an opposite sign to (\ref{wv-A}). 
\subsubsection{First-order equations for $dS_2\times \S_2$}
The equations of motion (\ref{dS2eom}) are solved by the same first-order equations (\ref{AdS2bps}) as in the $AdS_2$ case
\beq
\dot f&=& \frac{\sqrt{2}}{3} \, W\,\left(\frac{\lambda a e^{-2 g}}{g_1}+1\right),
\non
\dot g&=& \frac{\sqrt{2}}{3} \, W\, \left(1-\frac{\lambda a e^{-2 g}}{g_1}\right),\non
\dot \phi&=& -\frac{4\sqrt{2}}{3} \, \left(1-\frac{\lambda a e^{-2 g}}{g_1}\right) \,\frac{\partial W}{\partial \phi}, \label{dS2bps}
\eeq
 where $\lambda = 1$ for $\S_2 = S^2$ and $-1$ for $\S_2 = H^2$,
subject to the following constraint  between the gauge flux $a$ and gauge coupling constant $g_1$
\beq
a g_1 = \frac{1}{2},\label{dS2twistc}
\eeq
and the relation (\ref{vw-d})
\beq
V = -\lf[ \frac{8}{9}\lf(\frac{\pd W}{\pd \phi}\rr)^2-\frac{2}{3} W^2 \rr] \,\,.\nonumber
\eeq
Note that unless an explicit pseudo-superpotential is substituted in the first-order equations (\ref{dS2bps}), the relation (\ref{vw-d}) and the constraint (\ref{AdS2twist}) are not enough to solve the equations of motion (\ref{dS2eom}). 
\\\indent
Having established the first-order equations that solve the second-order field equations for both the $dS_4$ and $dS_2\times \S_2$ cases, we now move on to check whether there exists any pseudo-superpotential $W$ that satisfies the required relation (\ref{vw-d}) for both the type I and type II gauged theories. 
\subsubsection{Type I gauged theories}
 With the vector multiplet scalars truncated out, the scalar potential from the four type I gaugings that admit $dS_4$ solutions is given in (\ref{so3131d}),
\beq
V = -\frac{1}{2}g_1^2 e^{-\p}\lf(1-4 e^\p + e^{2\p}  \rr) \,\,.\nonumber
\eeq
This scalar potential appears in the equations of motion for both the cases of $dS_4$ (\ref{dS4eom}) and $dS_2\times \S_2$  (\ref{dS2eom}). 
The pseudo-superpotential that satisfies the relation (\ref{vw-d}) with $V$ given by (\ref{so3131d}) reads
\beq
W = -\frac{3i}{2\sqrt{2}}g_1 \, \lf[e^{\frac{\phi}{2}} - e^{\frac{-\phi}{2}} \rr]\,\,. \label{WdS2}
\eeq 
The first-order equations (\ref{dS4bps}) and (\ref{dS2bps}) with $W$ given in (\ref{WdS2})  solve the equations of motion (\ref{dS4eom}) and (\ref{dS2eom}), respectively. Although this is the case, these first-order equations do not give rise to either the $dS_4$ solution (\ref{so31-t1-dS4}) nor the comoslogical solutions interpolating between this $dS_4$ and the $dS_2\times\S_2$ fixed-point solutions (\ref{t1-so3131-dS2}). We elaborate more on this below. 
\ben
\item \textit{$dS_4$ solution}. The $dS_4$ solution (\ref{so31-t1-dS4}) admitted by  $V$ (\ref{so3131d}) 
\beq
\phi_0 =  0, \qquad f_0= \frac{g_1}{\sqrt{3}}, \nonumber
\eeq
is not a critical point of the pseudo-superpotential (\ref{WdS2}). 
This is because given (\ref{vw-d}), the extremization of $V$, $V'=0$, implies either one of the two following conditions
\beq
V' = 0  \rd  W'=0, \qquad \text{or}\qquad 3 W - 4W''=0 \,.\label{vw-dd}
\eeq
At the $dS_4$ point $\phi_0 = 0$, although $W'\neq 0$,
\beq
(3 W - 4 W'') =0,
\eeq
leading to $V'=0$. As such, the $dS_4$ solution (\ref{so31-t1-dS4}) cannot arise from the first-order equations (\ref{dS4bps}). It was pointed out in \cite{ST0610} that in general, for a cosmological solution, if a scalar potential can be written in terms of a pseudo-superpotential as
\beq
 V = -2 \lf(W'^2 - \a^2 W^2\rr),
\eeq
where $\a$ is a constant, then the critical point at which $V'=0$ can arise from either
\beq
W'=0 \qquad \text{or}\qquad  W''-\a^2 W=0.
\eeq 
When the former condition is satisfied then the solution is pseudosupersymmetric and $W$ is the corresponding pseudo-superpotential. When the latter condition is satisfied then the solution is not pseudo-supersymmetric. These conditions are tied to the Breitenlohner-Freedman (BF) bound of the solutions, which, in the case of cosmology, reads
\beq
m^2 \leq \frac{(D-1)^2}{4L^2}, \label{BFb}
\eeq
where $D$ is the spacetime dimension and $L=1/f_0$ is the radius of $dS$ space. Solutions admitted by the pseudo-superpotential $W$ (corresponding to the case $W'=0$) do not violate the BF bound, while those not admitted by the pseudo-superpotential $W$ (corresponding to the case $W''-\a^2 W=0$) do. This argument follows from the work of \cite{ST0610}. In this case, 
for $D=4$ and $L  = \sqrt{\frac{3}{V_0}}= 1/f_0$, it can be verified from \cite{deRoo-Panda2} that $dS_4$ solutions of the four gaugings in the type I theories do violate the BF bound. We explicitly list the mass values violating this BF bound for each of the four type I gauged theories in Table \ref{BF-bounds}.
\item \textit{$dS_2\times \S_2$ cosmological solutions}. 
Although the first-order equations (\ref{dS2bps}) with $W$ (\ref{WdS2}) and the second-order equations (\ref{dS2eom}) with $V$ (\ref{so3131d}) both admit the fixed-point solutions (\ref{t1-so3131-dS20}), the cosmological solutions interpolating between this fixed point solutions and the $dS_4$ solution (\ref{so31-t1-dS4}) as given in Fig.\ref{fig:so31-t1-dS2} are not  solutions of (\ref{dS2bps}). The reason for this is twofold. Firstly, the pseudo-superpotential (\ref{WdS2}) does not admit $dS_4$ solution (\ref{so31-t1-dS4}) as $V$ (\ref{so3131d}), as pointed out in the previous section. Secondly, the constraint (\ref{dS2twistc}) is needed so that solutions of the equations of motion (\ref{dS2eom}) become solutions of (\ref{dS2bps}). However, once (\ref{dS2twistc}) is imposed, $f_0$ in (\ref{t1-so3131-dS20}) vanishes (for both $\kappa = 1$ and $-1$)  which  renders the solution physically unacceptable. So, neither the $dS_4$ solution (\ref{so31-t1-dS4}) nor the fixed-point solutions (\ref{t1-so3131-dS2}) can be admitted by the first-order equations (\ref{dS2eom}). Instead,
the  fixed point solutions (\ref{t1-so3131-dS2}) that are real  and their associated cosmological solutions Fig.\ref{fig:so31-t1-dS2} can only arise from the second-order equations of motion (\ref{dS2eom})  with the constraint (\ref{dS2twist}). 
\enn
\subsubsection{Type II gauged theories}
For the scalar potentials given in Table \ref{dS4-sum-V}, no suitable $W$ can be found such that (\ref{vw-d}) is satisfied. Instead, we found the following $W$ 
\beq
W  = c\, g_1\, \lf(e^{\p/2}-e^{-\p/2} \rr),
\qquad c = \begin{dcases} 
\pm\frac{3}{2\sqrt{2}}, & \begin{array}{l} SO(2,1)\times SO(2,1) \end{array}\\ \\
\pm \frac{3}{2}, & \begin{array}{l}   SO(2,1)\times SO(2,2) \end{array}\\ \\
\pm\lf(\frac{3}{2}\rr)^{3/2}, &\begin{array}{l} SO(3,1)\times SO(2,1) \\ SO(3,1)\times SO(3,1)  \end{array}\\ \\
\pm\frac{3 \sqrt{3}}{2},
&\begin{array}{l}  SO(2,1)\times SO(4,1)\end{array}\\\\
\pm 3\sqrt{\frac{3}{2}}, &\begin{array}{l} SO(2,1)\times SU(2,1) \end{array}
 \end{dcases}
 \label{t2-W}
\eeq
that satisfies the following relation
\beq
V=\frac{8}{9}\lf(\frac{\partial W}{\partial \phi}\rr)^2+\frac{2}{9} W(\phi)^2. \label{vw-d2}
\eeq
Consequently, there do not exist any systems of first-order equations that solve the second-order equations (\ref{dS4eom}) for the $dS_4$ case or (\ref{dS2eom}) for the $dS_2\times \S_2$ case for the type II $dS$ solutions. 
It is worth recalling that since the type II $dS$ gauged theories in 4D are directly related to 5D $dS$ gauged theories via dimensional reduction as shown in \cite{dS4}, this situation agrees with the result from the five-dimensional analysis \cite{dS5-cosmo} which shows there do not exist any suitable pseudo-superpotentials (and systems of first-order equations) for the $dS_5$ and $dS_{2,3}\times \S^{3,2}$ cosmological solutions.
\\\indent
 We also note that the pseudo-superpotentials $W$ as given in (\ref{t2-W}) do not admit the $dS_4$ solutions at $\phi_0 = \chi_0 = 0$ that are admitted by the scalar potentials $V$ given in (\ref{dS4-sum-V}). The reason for this is the same as the type I case above. Given the relation (\ref{vw-d2}), the extremization of $V$, $V'=0$, can be obtained from either
\beq
W'=0, \qquad \text{or}\qquad (W+ 4W'')=0
\eeq
At the $dS_4$ point $\phi_0 = 0$, $W'\neq 0$, but rather
\beq
W +4 W'' = 0. 
\eeq
It can again be verified from \cite{deRoo-Panda2} and \cite{dS4} that $dS_4$ solutions of the six gaugings in the type II theories violate the BF bound (\ref{BFb}).  The mass values violating the BF bound (\ref{BFb}) for each of the six type II gauged theories  are listed in Table \ref{BF-bounds}. 
\subsubsection{BF bounds for $dS_4$ solutions}
For $D=4$, the BF bound (\ref{BFb}) reads
\beq
m^2 \leq \frac{9}{4} L^2,  \qquad\text{or}\qquad m^2 L^2 \leq \frac{9}{4}\label{BFb4d}
\eeq
 Given that the $dS_4$ length $L$ is related to the extremized value $V_0$ of $V$ as
\beq
L^2 = \frac{3}{V_0},
\eeq
the bound (\ref{BFb4d}) can also be written in terms of $V_0$ 
\beq
m^2 \leq \frac{3V_0}{4}\,\,. \label{BFv}
\eeq
 The mass spectra of $dS_4$ solutions in all $dS$ gauged theories, excluding $SO(2,1)\times SO(4,1)$, can be found in \cite{deRoo-Panda2}. The mass spectrum of the $dS_4$ solution of the $SO(2,1)\times SO(4,1)$ gauged theory can be found in \cite{dS4}. 
 In Table \ref{BF-bounds},  we list $V_0$ and the mass values violating the BF bound as given in \cite{deRoo-Panda2} and \cite{dS4}. Only one mass value for each gauged theory is sufficient to show that the BF bound is violated. Due to the different notations used in \cite{deRoo-Panda2} and \cite{dS4}, we will use (\ref{BFv}) for \cite{deRoo-Panda2} and (\ref{BFb4d}) for \cite{dS4}. 
 \\\indent
 Some comments regarding these notations are in order to avoid any potential confusion. In \cite{deRoo-Panda2}, $V_0$ is given in terms of $a_{ij}$ that is defined as
 \beq
 a_{ij} = g_i g_j\, \sin(\a_i - \a_j), \qquad i, j = 1, 2, \ldots
 \eeq
 where indices $i, j$ label the various individual gauge factors $G_1 \times G_2 \times \ldots $ constituting the gauge group $G_0$, 
  $g_i, g_j$ are the corresponding gauge couplings, and $\a_{i}$ are the $SU(1,1)\cong SL(2, \mb R)$ angles. Although there can be more than two gauge factors in $G_0$, all cases were shown to be reduced to just two factors (see Table \ref{dS4-sum-V}).
 The mass spectra are also given in terms of $a_{ij}$, making it convenient to check the BF bound using (\ref{BFv}). In \cite{dS4}, $V_0$ is given in terms of $g_1^2$ after applying the scaling ratio $g_2/g_1$ to bring the $dS_4$ critical point $\phi_0 \neq 0$ to the origin of the scalar manifold at $\phi_0=0$, while the mass spectrum is given in units of $m^2 L^2$. Accordingly, it is convenient to use (\ref{BFb4d}).
 \begin{table}[htb!]
 \centering
\begin{tabular}{|c|c|c|}
\hline 
 Gauge group $G_0$& $V_0$ & $\begin{array}{c}\text{Mass value $m^2$ violating (\ref{BFv})} \\ \lf(\times \,\text{multiplicities}\rr)\end{array}$
 \\
\hline&& \\
Type I &&\\
$SO(3)\times SO(3)$ &$\begin{array}{c}-|a_{12}| - 2a_{12}\\ a_{12}<0\end{array}$   & $\begin{array}{c} -2 a_{12} \,\,(\times \,36)
\end{array}$
\\&&\\
$SO(3,1)\times SO(3)$ &$\begin{array}{c}-|a_{12}| - 2a_{12}\\ a_{12}<0\end{array}$   & $\begin{array}{c} 2 |a_{12}| - 4 a_{12}\,\,(\times\, 5)
\end{array}$
\\&&\\
$SO(3,1)\times SO(3,1)$ &$\begin{array}{c}-|a_{12}| - 2a_{12}\\ a_{12}<0\end{array}$  & $\begin{array}{c} 2 |a_{12}| - 4 a_{12}\,\,(\times\,10)
\end{array}$
\\&&\\
$SL(3, \mb R)\times SO(3)$ &$\begin{array}{c}-|a_{12}| - 2a_{12}\\ a_{12}<0\end{array}$   & $\begin{array}{c} -2 a_{12} \,\,(\times\,6)
\end{array}$\\ \hline &&\\
Type II &&\\
$SO(2,1)\times SO(2,1)$ & $|a_{12}| $  & $\begin{array}{c}  |a_{12}|\,\,(\times\,4)
\end{array}$ \\&&\\
$SO(2,1)\times SO(2,2)$ & $\begin{array}{c}\sqrt{\Delta_0} =\sqrt{a_{12}^2 + a_{13}^2 +  a^2_{23}} \end{array}$ &     $\begin{array}{c}\dfrac{1}{\sqrt{\Delta_0}}(a_{12}^2 + a_{13}^2) + 2 a_{23}\\ a_{12} = a_{13} = a_{23} = \sqrt{\dfrac{\Delta_0}{3}} \\ (\times \,2)
\end{array}$\\&&\\
$SO(3,1)\times SO(2,1)$ & $\sqrt{3}|a_{12}|$  & $\begin{array}{c} \sqrt{3}|a_{12}\,\,(\times\,6)
\end{array}$\\ &&\\
$SO(3,1)\times SO(3,1)$ & $3|a_{12}| $   & $\begin{array}{c} 8|a_{12}|\,\,(\times\,1)
\end{array}$\\&&\\
$SU(2,1)\times SO(2,1)$ &$2\sqrt{3} |a_{12}|$ & $\begin{array}{c} 2\sqrt{3}|a_{12}|\,\,(\times\,10)
\end{array}$\\&&\\
$SO(2,1)\times SO(4,1)$ & & $\begin{array}{c}\text{BF bound is (\ref{BFb4d})}\\ m^2 L^2 = 6 \,\,(\times 2) \end{array}$
\\ &&\\\hline
\end{tabular}
\caption{Mass values (and their multiplicities) violating the BF bound (\ref{BFv}) for the 10 $dS$ gauged theories taken from \cite{deRoo-Panda2} with the exception of the gauge group $SO(2,1)\times SO(4,1)$ which is taken from \cite{dS4}.} \label{BF-bounds}
\end{table}  
\clearpage
\subsection{Summary}
For both types of $dS$ gauged theories,
we checked whether the first-order systems of equations, to which the second-order field equations reduce, can give rise to the $dS_4$ and cosmological solutions of sections \ref{4D-dS2-t1}, \ref{4D-dS2-t2}. Following the domain wall/cosmology correspondence established in \cite{ST0610}, this analysis was performed using the $AdS$ case as a reference, since the equations of motion for the $AdS$ case are almost identical to those of the $dS$ case, except for a few opposite signs in front of some non-derivative terms. The first-order equations for the $dS_4$ and $dS_2\times \S_2$ solutions, if they exist, should be identical in form to the BPS equations for $AdS_4$ and $AdS_2\times \S_2$ solutions, respectively. The reducibility of the second-order field equations to the first-order equations hinges on there being a required relation $V(W)$ between $V$ and $W$ such that $V$ can be written in terms of the squares of $W$ and its derivatives.
This $V(W)$ relation in the $dS$ case should be identical, save for an overall opposite sign, to the $V(W)$ relation in the $AdS$ case. While supersymmetry automatically ensures the existence of the superpotential $W$ in the $AdS$ case, there is no guarantee that a suitable pseudo-superpotential $W$ can be found in the $dS$ case. Thus, while the general form of the first-order equations in the $dS$ case is settled, its validity is only confirmed if a suitable $W$ exists. The suitability of $W$ is decided by two conditions, the first one being whether it satisfies the required $V(W)$ relation, and the second one being whether it admits the same $dS$ critical point as $V$. 
\\\indent
 For the type I $dS$ gauged theories,  we did find a pseudo-superpotential $W$ satisfying the required $V(W)$ relation, but this $W$ does not admit the $dS_4$ solution that is admitted by the scalar potential $V$. Hence, the first suitability condition is satisfied but the second is not. This eliminates the possibility of the type I $dS_4$ solution and its associated $dS_2\times \S_2$ cosmological solutions arising from the first-order equations that have the same form as the BPS equations in the $AdS$ case. 
For the type II $dS$ gauged theories, on the other hand, there does not exist any $W$ that satisfies the required $V(W)$ relation. Thus, even the first suitability condition fails. This is almost the same as the result of a similar analysis in the five-dimensional case \cite{dS5-cosmo} where there is only one type of $dS$ gaugings that corresponds to type II gaugings in four dimensions. In five dimensions, although the first suitability condition is not fulfilled in an exact manner as in the type II theories in four dimensions, the second one is. Regardless of the exact way the pseudo-superpotential fails to be suitable in either the 4D type I or type II theories, and even in the 5D $dS$ gauged theories, this failure can be traced to the fact that firstly these $dS$ solutions are unstable, and secondly they violate the BF bound that serves as a means to guarantee pseudo-supersymmetry.
\section{Concluding remarks}\label{concl}
In this work, we have studied cosmological solutions interpolating between a $dS_2\times \S_2$ spacetime, with $\S_2 = S^2$ and $H^2$,  and a $dS_4$ spacetime from $N=4$ four-dimensional gauged supergravity. We emphasize that our motivation for the study of these solutions is completely decoupled from any holographic contexts. Instead, we are solely motivated by the question of whether cosmological solutions exist in 4D $N=4$ supergravity, given the existence of $dS_4$ vacua \cite{dS4}. Consequently, the cosmological solutions found were obtained by solving the second-order field equations in theories with gauge groups capable of admitting $dS_4$ solutions.
Although the methodology used in this work is the same as that used in the 5D work \cite{dS5-cosmo}, there is an important distinction between the two. In 5D there is only one type of gaugings that can admit $dS$ vacua.
In 4D there are two types of gaugings. 
Type I $dS$ gauged theories consist of four theories with gauge groups $SO(3)\times SO(3)$, $SO(3,1)\times SO(3,1)$, $SO(3)\times SO(3,1)$, $SO(3)\times SL(3,\mb R)$ whose common compact subgroup $SO(3)\times SO(3)$ is fully embedded in the $SO(6)$ R-symmetry directions. These type I theories can admit both $dS_4$ and $AdS_4$ solutions. The cosmological solutions in the type I theories require an Abelian vector field that corresponds to the diagonal $U(1)$ of the $U(1)\times U(1)$ subgroup of the aforementioned $SO(3)\times SO(3)$ compact part.
 Type II $dS$ gauged theories consist of six theories with gauge groups $SO(2,1)\times SO(2,1)$, $SO(2,1)\times SO(2,2)$, $SO(2,1)\times SO(3,1)$, $SO(3,1)\times SO(3,1)$, $SO(2,1)\times SO(4,1)$ and $SO(2,1)\times SU(2,1)$ whose compact subgroups are entirely embedded in the $SO(n)$ matter symmetry directions. These type II theories can only admit $dS_4$ solutions without the possibility of admitting $AdS_4$ solutions. To obtain cosmological solutions in the type II $dS$ theories, we needed to turn on an Abelian gauge field corresponding to the diagonal of the $U(1)\times U(1)$ subgroup of the compact part along the $SO(n)$ matter directions. Only
 4D type II gaugings can be derived from 5D $dS$ gaugings, while 4D type I gaugings can be derived from 5D $AdS$ gaugings \cite{dS4}. This crucial difference between 5D and 4D theories is an important motivation to explore the 4D theories given the results of the 5D case \cite{dS5-cosmo}. 
 \\\indent
 Furthermore, we also characterized the extent to which these nonsupersymmetric $dS_4$ and their associated cosmological solutions fail to arise from the relevant first-order equations that solve the second-order field equations by studying the lack of suitable pseudosuperpotentials in the type I and II $dS$ gauged theories. 
 Finally, we note that cosmological solutions arising from either type I or type II theories require the square of the product of the gauge flux $a$ and gauge coupling $g_1$ to be negative in order to be real. This feature is already encountered in the five-dimensional cosmological solutions \cite{dS5-cosmo} and resembles the situation in the $dS$ supergravities, arising from the dimensional reduction of M$\star$/IIB$\star$-theories, with the wrong sign for the gauge kinetic terms. 
 Instead of regarding these solutions as pathological due to this particular feature, we interpret this as an additional characterization of the $dS$ vacuum structure of half-maximal supergravity. The implications of this remain to be understood and we hope that more work will elucidate this matter further. 
\\\\
\textbf{Acknowledgements}: HLD is supported by the grants C‐144‐000‐207‐532 and C‐141‐000‐777‐532 for postdoctoral research. 

\end{document}